\documentclass[preprint,english,aps,prl,showpacs,preprintnumbers,amsmath,amssymb,superscriptaddress]{revtex4}
\usepackage[latin9]{inputenc}
\usepackage{amsmath}
\usepackage{graphicx}

\makeatletter
\@ifundefined{textcolor}{}
{%
 \definecolor{BLACK}{gray}{0}
 \definecolor{WHITE}{gray}{1}
 \definecolor{RED}{rgb}{1,0,0}
 \definecolor{GREEN}{rgb}{0,1,0}
 \definecolor{BLUE}{rgb}{0,0,1}
 \definecolor{CYAN}{cmyk}{1,0,0,0}
 \definecolor{MAGENTA}{cmyk}{0,1,0,0}
 \definecolor{YELLOW}{cmyk}{0,0,1,0}
}

\usepackage{dcolumn}
\usepackage{bm}
\usepackage{color}

\makeatother

\usepackage{babel}
\begin{document}

\title{Direct evidence for the emergence of a pressure induced nodal superconducting gap in the iron-based superconductor Ba$_{0.65}$Rb$_{0.35}$Fe$_{2}$As$_{2}$}

\author{Z.~Guguchia}

\email{zurab.guguchia@psi.ch}

\selectlanguage{english}%

\affiliation{Laboratory for Muon Spin Spectroscopy, Paul Scherrer Institute, CH-5232
Villigen PSI, Switzerland}

\author{A.~Amato}

\affiliation{Laboratory for Muon Spin Spectroscopy, Paul Scherrer Institute, CH-5232
Villigen PSI, Switzerland}

\author{J.~Kang}

\affiliation{School of Physics \& Astronomy, University of Minnesota, Minneapolis,
MN 55455, USA}

\author{H.~Luetkens}

\affiliation{Laboratory for Muon Spin Spectroscopy, Paul Scherrer Institute, CH-5232
Villigen PSI, Switzerland}

\author{P.K.~Biswas}

\affiliation{Laboratory for Muon Spin Spectroscopy, Paul Scherrer Institute, CH-5232
Villigen PSI, Switzerland}

\author{G.~Prando}

\affiliation{Leibniz-Institut f\"{u}r Festk\"{o}rper- und Werkstoffforschung (IFW) Dresden,
D-01171 Dresden, Germany}

\author{ F.~von Rohr}

\affiliation{Physik-Institut der Universit\"{a}t Z\"{u}rich, Winterthurerstrasse 190,
CH-8057 Z\"{u}rich, Switzerland}

\author{Z.~Bukowski}

\affiliation{Institute of Low Temperature and Structure Research, Polish Academy
of Sciences, 50-422 Wroclaw, Poland}

\author{A.~Shengelaya}

\affiliation{Department of Physics, Tbilisi State University, Chavchavadze 3,
GE-0128 Tbilisi, Georgia}

\author{H.~Keller}

\affiliation{Physik-Institut der Universit\"{a}t Z\"{u}rich, Winterthurerstrasse 190,
CH-8057 Z\"{u}rich, Switzerland}

\author{E.~Morenzoni}

\affiliation{Laboratory for Muon Spin Spectroscopy, Paul Scherrer Institute, CH-5232
Villigen PSI, Switzerland}

\author{R.M. Fernandes}

\affiliation{School of Physics \& Astronomy, University of Minnesota, Minneapolis,
MN 55455, USA}

\author{R.~Khasanov}

\affiliation{Laboratory for Muon Spin Spectroscopy, Paul Scherrer Institute, CH-5232
Villigen PSI, Switzerland}
\begin{abstract}
Identifying the superconducting (SC) gap structure of the iron-based
high-temperature superconductors (Fe-HTS's) remains a key issue for the understanding of superconductivity in these materials. 
In contrast to other unconventional superconductors, in
the Fe-HTS's both $d$-wave and extended $s$-wave pairing symmetries
are close in energy, with the latter believed to be generally favored
over the former. Probing the proximity between these very different
SC states and identifying experimental parameters that can tune them,
are of central interest. Here we report high-pressure muon spin rotation
experiments on the temperature-dependent magnetic penetration depth
$\lambda\left(T\right)$ in the optimally doped Fe-HTS Ba$_{0.65}$Rb$_{0.35}$Fe$_{2}$As$_{2}$.
At ambient pressure this material is known to be a nodeless $s$-wave
superconductor. Upon pressure a strong decrease of $\lambda\left(0\right)$
is observed, while the SC transition temperature remains nearly constant.
More importantly, the low-temperature behavior of $1/\lambda^{2}\left(T\right)$
changes from exponential saturation at zero pressure to a power-law
with increasing pressure, providing unambiguous evidence that hydrostatic
pressure promotes nodal SC gaps. Comparison to microscopic models
 favors a $d$-wave over a nodal $s^{+-}$-wave pairing as the origin
of the nodes. Our results provide a new route of understanding the
complex topology of the SC gap in Fe-HTS's. 
\end{abstract}


\maketitle
\section{INTRODUCTION}
After six years of intensive research on the Fe-based high temperature
superconductors (Fe-HTS's), no consensus on a universal gap structure
has been reached. There is evidence that small differences in electronic
or structural properties can lead to a strong diversity in the superconducting
(SC) gap structure. On the one hand, nodeless isotropic gap functions
were observed in optimally doped Ba$_{1-x}$K$_{x}$Fe$_{2}$As$_{2}$,
Ba$_{1-x}$Rb$_{x}$Fe$_{2}$As$_{2}$ and BaFe$_{2-x}$Ni$_{x}$As$_{2}$
as well as in BaFe$_{2-x}$Co$_{x}$As$_{2}$, K$_{x}$Fe$_{2-y}$Se$_{2}$,
and FeTe$_{1-x}$Se$_{x}$ \cite{Ding,KhasanovN,GuguchiaN,TerashimaN,ZhangY,MiaoN,Abdel-Hafiez,BiswasPRB}.
On the other hand, signatures of nodal SC gaps were reported in LaOFeP,
LiFeP, KFe$_{2}$As$_{2}$, BaFe$_{2}$(As$_{1-x}$P$_{x}$)$_{2}$,
BaFe$_{2-x}$Ru$_{x}$As$_{2}$, FeSe as well as in over-doped Ba$_{1-x}$K$_{x}$Fe$_{2}$As$_{2}$
and BaFe$_{2-x}$Ni$_{x}$As$_{2}$ \cite{Abdel-Hafiez,FlechterN,HashimotoN,YamashitaN,NakaiN,HashimotoK,DongN,QiuN,SongN,ZhangN}.
Understanding what parameters of the systems control the different
SC gap structures observed experimentally is paramount to elucidate
the microscopic pairing mechanism in the Fe-HTS's and, more generally,
to provide a deeper understanding of the phenomenon of high-temperature
superconductivity. On the theoretical front, it has been proposed
that both the $s^{+-}$-wave and $d$-wave states are close competitors
for the SC ground state \cite{Kuroki,Graser10,Thomale11,Maiti11,Khodas12,Fernandes13,FernandezN}.
Although the former generally wins, it has been pointed out that a
$d$-wave state may be realized upon removing electron or hole pockets.
On the experimental front, a sub-leading $d$-wave collective mode
was observed by Raman experiments inside the fully gapped SC state
of optimally doped Ba$_{1-x}$K$_{x}$Fe$_{2}$As$_{2}$ \cite{raman_mode,Bohm}.
In KFe$_{2}$As$_{2}$, a change of the SC pairing symmetry by hydrostatic
pressure has been recently proposed, based on the $V$-shaped pressure
dependence of $T_{{\rm c}}$ \cite{Taillefer}. However, no direct
experimental evidence for a pressure induced change of either the
SC gap symmetry or the SC gap structure in the Fe-HTS's has been reported until now.
Here, we show unambiguous evidence for the appearance of SC nodes
in optimally-doped Ba$_{1-x}$Rb$_{x}$Fe$_{2}$As$_{2}$ upon applied
pressure, consistent with a change from a nodeless $s^{+-}$-wave state to either
a nodal $s^{+-}$-wave or a $d$-wave state. 

Our results rely on measurements of the magnetic penetration depth
$\lambda$, which is one of the fundamental parameters of a superconductor,
since it is related to the superfluid density $n_{s}$ via 1/${\lambda}^{2}$
= $\mu_{0}$$e^{2}$$n_{s}/m^{*}$ (where $m^{*}$ is the effective
mass). Most importantly, the temperature dependence of ${\lambda}$
is particularly sensitive to the presence of SC nodes: while in a
fully gapped SC $\Delta\lambda^{-2}\left(T\right)\equiv\lambda^{-2}\left(0\right)-\lambda^{-2}\left(T\right)$
vanishes exponentially at low $T$, in a nodal SC it vanishes as a
power of $T$. The muon-spin rotation (${\mu}$SR) technique provides
a powerful tool to measure ${\lambda}$ in type II superconductors
\cite{Sonier}. A ${\mu}$SR experiment in the vortex state of a type
II superconductor allows the determination of ${\lambda}$ in the
bulk of the sample, in contrast to many techniques that probe ${\lambda}$
only near the surface.

For the compound Ba$_{0.65}$Rb$_{0.35}$Fe$_{2}$As$_{2}$ investigated
here, and for the closely related system Ba$_{1-x}$K$_{x}$Fe$_{2}$As$_{2}$,
previous $\mu$SR measurements of $\lambda\left(T\right)$ revealed
a nodeless multi-gap SC state \cite{KhasanovN,GuguchiaN}, in agreement
with ARPES measurements \cite{Ding,Evtushinsky,Zabolotnyy}. In this
communication, we report on ${\mu}$SR studies of ${\lambda}\left(0\right)$
and of the temperature dependence of $\Delta\lambda^{-2}$ in optimally
doped Ba$_{0.65}$Rb$_{0.35}$Fe$_{2}$As$_{2}$ under hydrostatic
pressures. This system exhibits the highest $T_{{\rm c}}$ ${\simeq}$
37 K among the extensively studied ``122'' family of Fe-HTS's. We
observe that while $T_{c}$ stays nearly constant upon application
of pressure, ${\lambda}\left(0\right)$ decreases substantially. In
view of previous works in another ``122'' compound that reported
a sharp peak of $\lambda(0)$ at a quantum critical point \cite{Hashimoto-Science},
we interpret the observed suppression of $\lambda(0)$ as evidence
that pressure moves the system away from a putative quantum critical
point in Ba$_{0.65}$Rb$_{0.35}$Fe$_{2}$As$_{2}$. More importantly,
we find a qualitative change in the low-temperature behavior of $\Delta\lambda^{-2}\left(T\right)$
as pressure is increased. While at $p=0$ an exponential suppression
characteristic of a nodeless superconductivity is observed, for $p=2.25$
GPa a clear power-law behavior is found. 
Because pressure does not affect the impurity concentration, which
could promote power-law behavior even for a nodeless system \cite{Hisrchfeld13},
our findings provide strong evidence for a nodeless to nodal SC transition.
Our fittings to microscopic models reveal that this behavior is more
compatible with a $d$-wave state rather than an $s^{+-}$ state with
accidental nodes, indicating that pressure promotes a change in the
pairing symmetry.

\section{RESULTS}

\subsection{Probing the nonuniform field distribution in the vortex state under
pressure}

\begin{figure}[t!]
\includegraphics[width=1\linewidth]{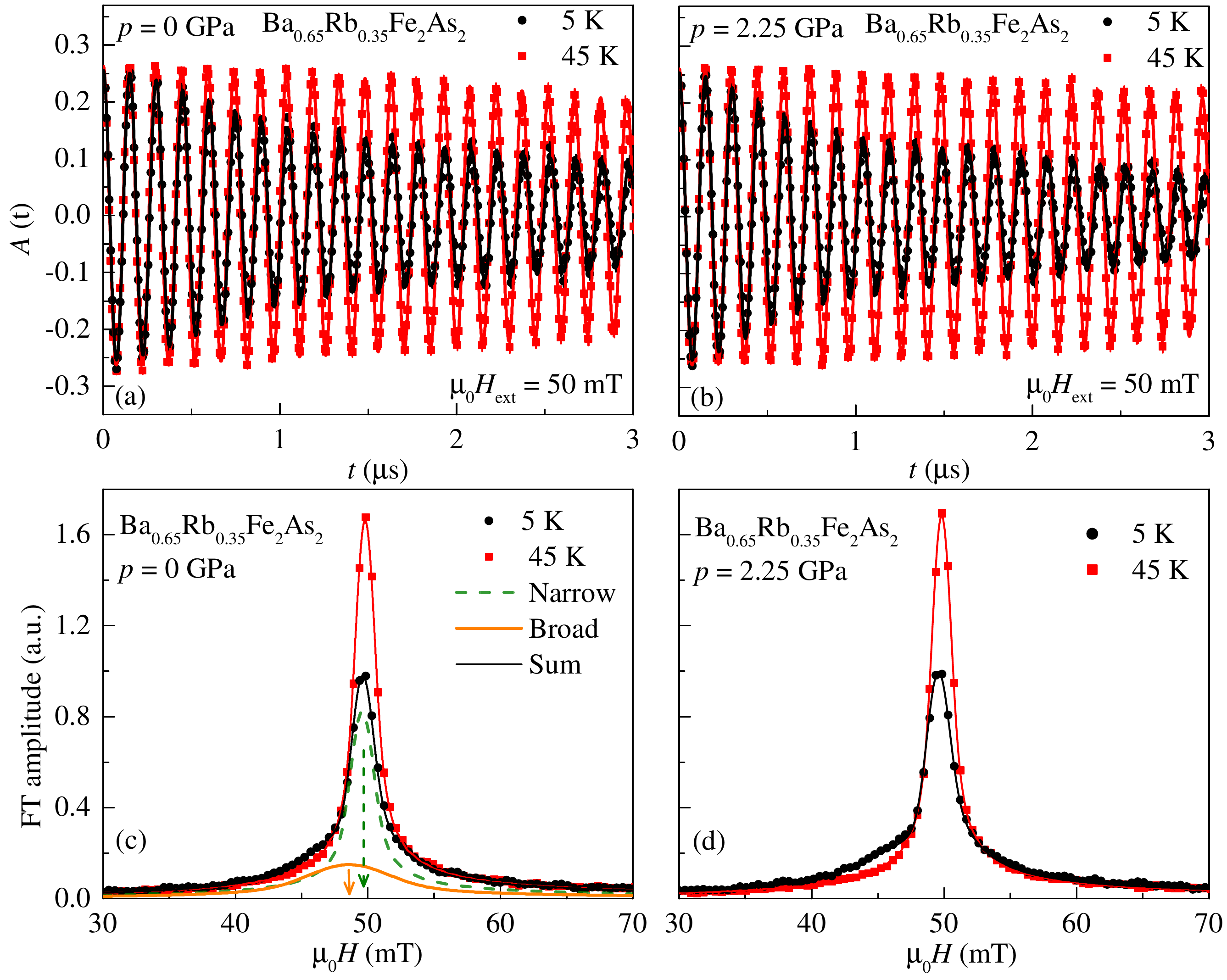} \protect\protect\caption{(Color online) \textbf{Transverse-field (TF) $\mu$SR time spectra
and the corresponding Fourier transforms (FT's) of Ba$_{0.65}$Rb$_{0.35}$Fe$_{2}$As$_{2}$.}
The spectra are obtained above (45 K) and below (5 K) $T_{{\rm c}}$
(after field cooling the sample from above $T_{{\rm c}}$): (a,c)
$p$ = 0 GPa and (b,d) $p$ = 2.22 GPa. The solid lines in panels
a and b represent fits to the data by means of Eq.~3. The solid lines
in panels c and d are the FT's of the fitted time spectra. The dashed
and solid arrows indicate the first moments for the signals of the
pressure cell and the sample, respectively.}

\label{fig1} 
\end{figure}

\begin{figure}[t!]
\includegraphics[width=0.6\linewidth]{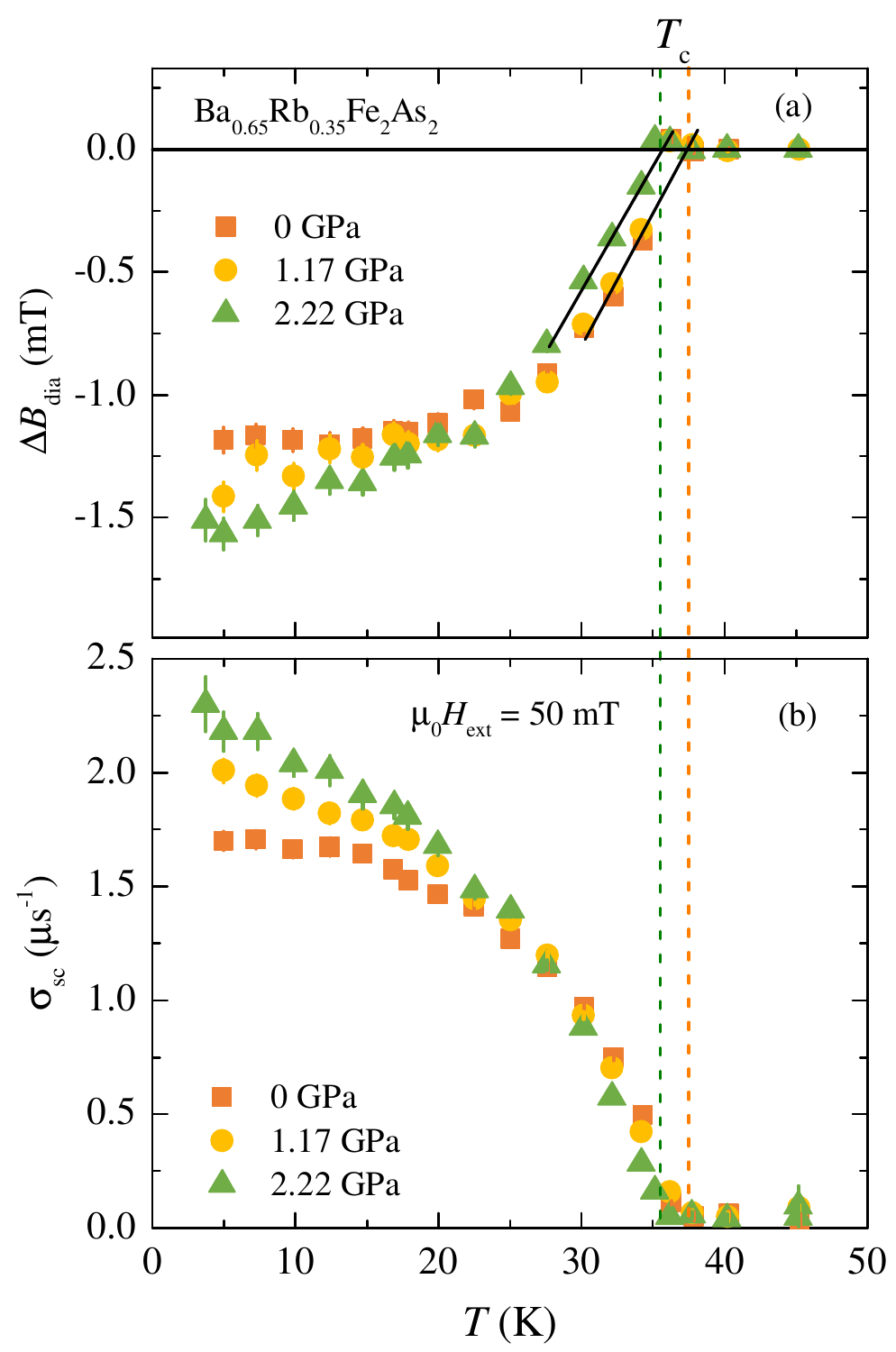} \vspace{-0.5cm}
 \protect\protect\caption{(Color online) \textbf{Diamagnetic shift ${\Delta}$$B_{{\rm dia}}$
(a) and the muon spin relaxation rate ${\sigma}_{{\rm sc}}$ (b) of
Ba$_{0.65}$Rb$_{0.35}$Fe$_{2}$As$_{2}$ as a function of temperature
at various pressures.} (a) The definition of the diamagnetic shift
${\Delta}$$B_{{\rm dia}}$ is given in the text. (b) The muon spin
relaxation rate ${\sigma}_{{\rm sc}}$ is measured in a magnetic field
of ${\mu_{0}}H$ = 50 mT. The dashed vertical lines denote $T_{c}$
for $p$ = 0 and 2.22 GPa.}

\label{fig7} 
\end{figure}

Figures 1a and b exhibit the transverse-field
(TF) ${\mu}$SR-time spectra for Ba$_{0.65}$Rb$_{0.35}$Fe$_{2}$As$_{2}$,
measured at ambient $p$ = 0 GPa and maximum applied pressure $p$
= 2.25 GPa, respectively. The spectra above (45 K) and below (1.7
K) the SC transition temperature $T_{{\rm c}}$ are shown. Above $T_{{\rm c}}$
the oscillations show a small relaxation due to the random local fields
from the nuclear magnetic moments. Below $T_{{\rm c}}$ the relaxation
rate strongly increases with decreasing temperature due to the presence
of a nonuniform local magnetic field distribution as a result of the
formation of a flux-line lattice (FLL) in the SC state. Figures 1c
and d show the Fourier transforms (FT's) of the ${\mu}$SR time spectra
shown in Figs.~1a and b, respectively. At $T$ = 5 K the narrow signal
around ${\mu}$$_{{\rm 0}}$$H_{\mathrm{ext}}$ = 50 mT (see Figs.~1c
and d) originates from the pressure cell, while the broad signal with
a first moment $\mu_{0}H_{\mathrm{int}}<\mu_{0}H_{\mathrm{ext}}$,
marked by the solid arrow in Fig.~1c, arises from the SC sample.

\begin{figure*}[ht!]
\centering \includegraphics[width=1.0\linewidth]{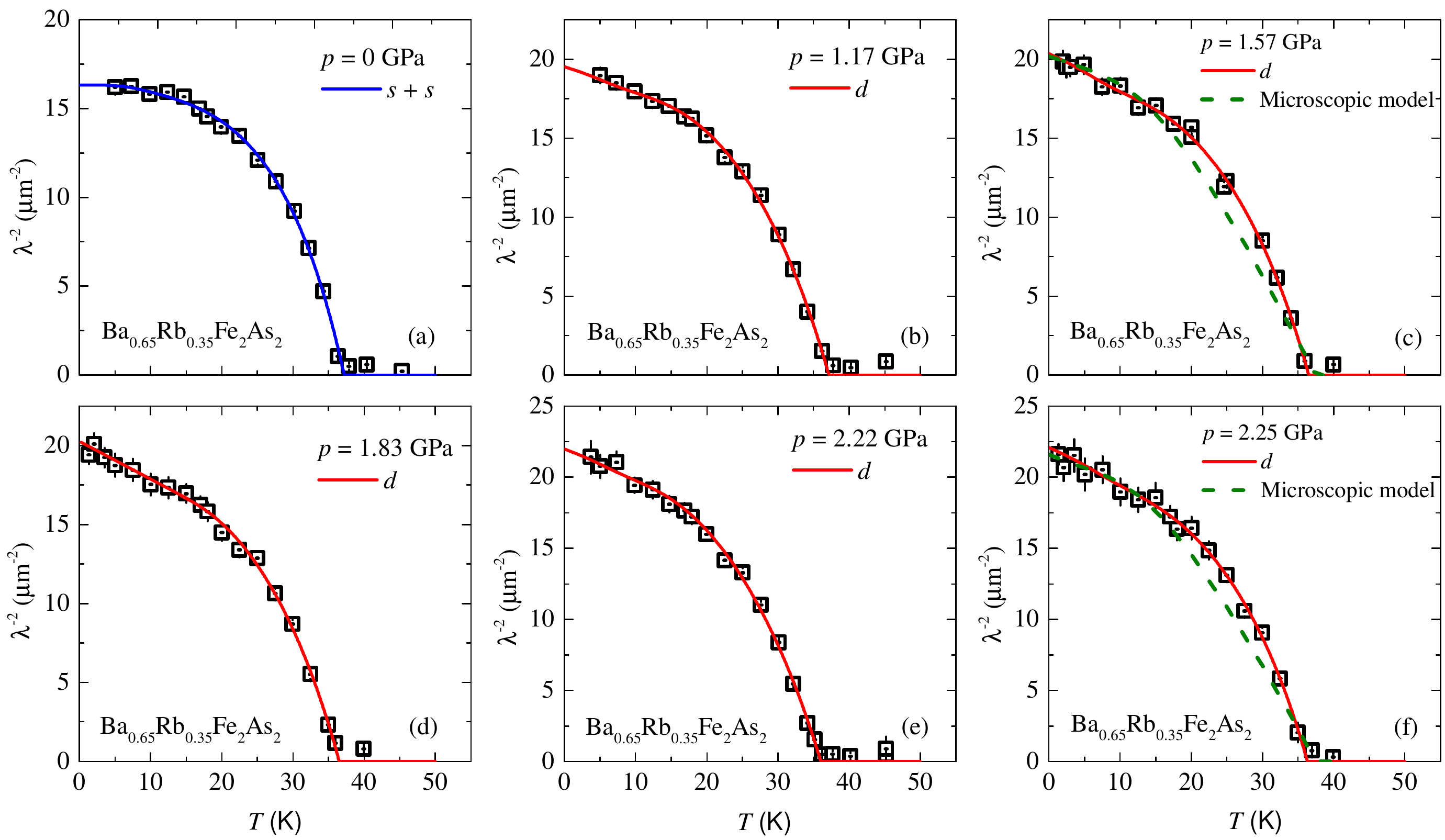} \vspace{-0.8cm}
 \protect\protect\caption{(Color online) \textbf{Pressure evolution of ${\lambda}^{-2}(T)$
of Ba$_{0.65}$Rb$_{0.35}$Fe$_{2}$As$_{2}$.} The temperature dependence
of ${\lambda}^{-2}$ measured at various applied hydrostatic pressures
for Ba$_{0.65}$Rb$_{0.35}$Fe$_{2}$As$_{2}$. The solid line for
$p$ = 0 GPa corresponds to a two-gap $s$-wave model (a) and the
solid lines for finite pressure represent a fits to the data using
a multiband $d$-wave model (b-f). The dased lines in panels (c) and
(f) represent fits to the data using the microscopic model described
in the supplemental material. }

\label{fig3} 
\end{figure*}


Below $T_{{\rm c}}$ a large diamagnetic shift of ${\mu}_{{\rm 0}}$$H_{{\rm int}}$
experienced by the muons is observed at all applied pressures. This
is evident in Fig.~2a, where we plot the temperature dependence of
the diamagnetic shift ${\Delta}$$B_{{\rm dia}}$ = ${\mu}_{{\rm 0}}${[}$H_{{\rm int,SC}}$-$H_{{\rm int,NS}}${]}
for Ba$_{0.65}$Rb$_{0.35}$Fe$_{2}$As$_{2}$ at various pressures,
where ${\mu}_{{\rm 0}}$$H_{{\rm int,SC}}$ denotes the internal field
measured in the SC state and ${\mu}_{{\rm 0}}$$H_{{\rm int,NS}}$
the internal field measured in the normal state at 45 K. Note, that ${\mu}_{{\rm 0}}$$H_{{\rm int,NS}}$ is temperature independent. This diamagnetic shift indicates the bulk character of superconductivity and excludes
the possibility of field induced magnetism \cite{Sonier2011} in Ba$_{0.65}$Rb$_{0.35}$Fe$_{2}$As$_{2}$
at all applied pressures. The SC transition temperature $T_{{\rm c}}$
is determined from the intercept of the linearly extrapolated ${\Delta}$$B_{{\rm dia}}$
curve its zero line (we used the same criterium for determination of $T_{{\rm c}}$ from ${\Delta}$$B_{{\rm dia}}(T)$ as from the susceptibility data ${\chi}_{\rm m}(T)$, presented in the supplementary material). It is found to be $T_{{\rm c}}$ = 36.9(7) K and 35.9(5) K
for $p$ = 0 GPa and 2.25 GPa, respectively. The ambient pressure
value of $T_{{\rm c}}$ is in perfect agreement with $T_{{\rm c}}$
= 36.8(5) K obtained from susceptibility and specific heat measurements
(see supplemental material). At the highest pressure of $p$ = 2.25
GPa applied, $T_{{\rm c}}$ decreases only by ${\simeq}$ 1 K, indicating
only a small pressure effect on $T_{{\rm c}}$ in Ba$_{0.65}$Rb$_{0.35}$Fe$_{2}$As$_{2}$.
The temperature dependence of the muon spin depolarization rate ${\sigma}_{{\rm sc}}$,
which is proportional to the second moment of the field distribution (the moments of the field distribution probed by the muons were extracted
with the equations described in the Method section), of Ba$_{0.65}$Rb$_{0.35}$Fe$_{2}$As$_{2}$ in the SC state at selected pressures is shown in Fig.~2b. 
Below $T_{{\rm c}}$ the relaxation rate ${\sigma}_{{\rm sc}}$ starts to
increase from zero with decreasing temperature due to the formation
of the FLL. It is interesting that the low-temperature value ${\sigma}_{{\rm sc}}$(5
K) increases substantially under pressure (see Fig.~2b): ${\sigma}_{{\rm sc}}$(5
K) increases about 30 ${\%}$ from $p$ = 0 GPa to p = 2.25 GPa. Interestingly,
the form of the temperature dependence of ${\sigma}_{{\rm sc}}$,
which reflects the topology of the SC gap, changes as a function of
pressure. The most striking change is in the low-temperature behaviour
of ${\sigma}_{{\rm sc}}(T)$. At ambient pressure ${\sigma}_{{\rm sc}}(T)$
shows a flat behavior below $T$/$T_{{\rm c}}$ ${\simeq}$ 0.4, whereas
the high-pressure data exhibit a steeper (linear) temperature dependence
of ${\sigma}_{{\rm sc}}$($T$) below $T$/$T_{{\rm c}}$ ${\simeq}$
0.4. We show in the following how these behaviors indicate the appearance of nodes
in the gap function.

\subsection{Temperature and pressure dependent magnetic penetration depth}

In order to investigate a possible change of the symmetry of the SC
gap, we note that ${\lambda}(T)$ is related to the relaxation rate
${\sigma}_{{\rm sc}}(T)$ by the equation \cite{Brandt}: 
\begin{equation}
\frac{\sigma_{sc}(T)}{\gamma_{\mu}}=0.06091\frac{\Phi_{0}}{\lambda^{2}(T)},
\end{equation}
where ${\gamma_{\mu}}$ is the gyromagnetic ratio of the muon, and
${\Phi}_{{\rm 0}}$ is the magnetic-flux quantum. Thus, the flat $T$-dependence
of ${\sigma}_{{\rm sc}}$ observed at $p=0$ for low temperatures
(see Fig.~2b) is consistent with a nodeless superconductor, in which
$\lambda^{-2}\left(T\right)$ reaches its zero-temperature value exponentially.
On the other hand, the linear $T$-dependence of ${\sigma}_{{\rm sc}}$
observed at $p=2.25$ GPa (see Fig.~2b) indicates that $\lambda^{-2}\left(T\right)$
reaches $\lambda^{-2}(0)$ linearly which is characteristic of line
nodes. This is the most striking result of this communication: Pressure in an optimally-doped
Fe-HTS can tune a nodeless gap into a nodal gap. Although this qualitative
analysis is robust, it does not elucidate whether these nodes arise
due to a nodal $s^{+-}$ state or a $d$-wave state.

To proceed with a quantitative analysis, we consider the local (London)
approximation (${\lambda}$ ${\gg}$ ${\xi}$, where ${\xi}$ is the
coherence length) and first employ the empirical ${\alpha}$-model.
The latter, widely used in previous investigations of the penetration
depth of multi-band superconductors \cite{Bastian,Tinkham,carrington,padamsee,Fang,GuguchiaN,khasanovalpha},
assumes that the gaps occuring in different bands, besides a common
$T_{{\rm c}}$, are independent of each other. Then, the superfluid
density is calculated for each component separately \cite{GuguchiaN}
and added together with a weighting factor. For our purposes, a two-band
model suffices, yielding: 
\begin{equation}
\frac{\lambda^{-2}(T)}{\lambda^{-2}(0)}=\omega_{1}\frac{\lambda^{-2}(T,\Delta_{0,1})}{\lambda^{-2}(0,\Delta_{0,1})}+\omega_{2}\frac{\lambda^{-2}(T,\Delta_{0,2})}{\lambda^{-2}(0,\Delta_{0,2})},
\end{equation}
where ${\lambda}(0)$ is the penetration depth at zero temperature,
${\Delta_{0,i}}$ is the value of the $i$-th SC gap ($i=1$, 2) at
$T=0$~K, and ${\omega}_{i}$ is the weighting factor which measures
their relative contributions to ${\lambda^{-2}}$ (i.e. ${\omega}_{1}+{\omega}_{2}=1$).

The results of this analysis are presented in Figs.~3a-f, where the
temperature dependence of ${\lambda^{-2}}$ for Ba$_{0.65}$Rb$_{0.35}$Fe$_{2}$As$_{2}$
is plotted at various pressures. We consider two different possibilities
for the gap functions: either a constant gap, $\Delta_{0,i}=\Delta_{i}$,
or an angle-dependent gap of the form $\Delta_{0,i}=\Delta_{i}\cos2\varphi$,
where $\varphi$ is the polar angle around the Fermi surface. The
data at $p$ = 0 GPa are described remarkably well by two constant
gaps, ${\Delta}_{1}$ = 2.7(5) meV and ${\Delta}_{2}$ = 8.4(3) meV.
These values are in perfect agreement with our previous results \cite{GuguchiaN}
and also with ARPES experiments \cite{Evtushinsky}, pointing out
that most Fe-based HTS's exhibit two-gap behavior, characterized by
one large gap with $2{\Delta}_{2}/k_{{\rm B}}T_{{\rm c}}=7(2)$ and
one small gap with $2{\Delta}_{1}/k_{{\rm B}}T_{{\rm c}}=2.5(1.5)$.
In contrast to the case $p$ = 0 GPa, for all applied pressures ${\lambda^{-2}}(T)$
is better described by one constant gap and one angle-dependent gap,
confirming the presence of gap nodes, as inferred from our qualitative
analysis. Note that a fitting to two angle-dependent gaps is inconsistent
with the data.

\begin{figure}
\begin{centering}
\includegraphics[width=0.6\columnwidth]{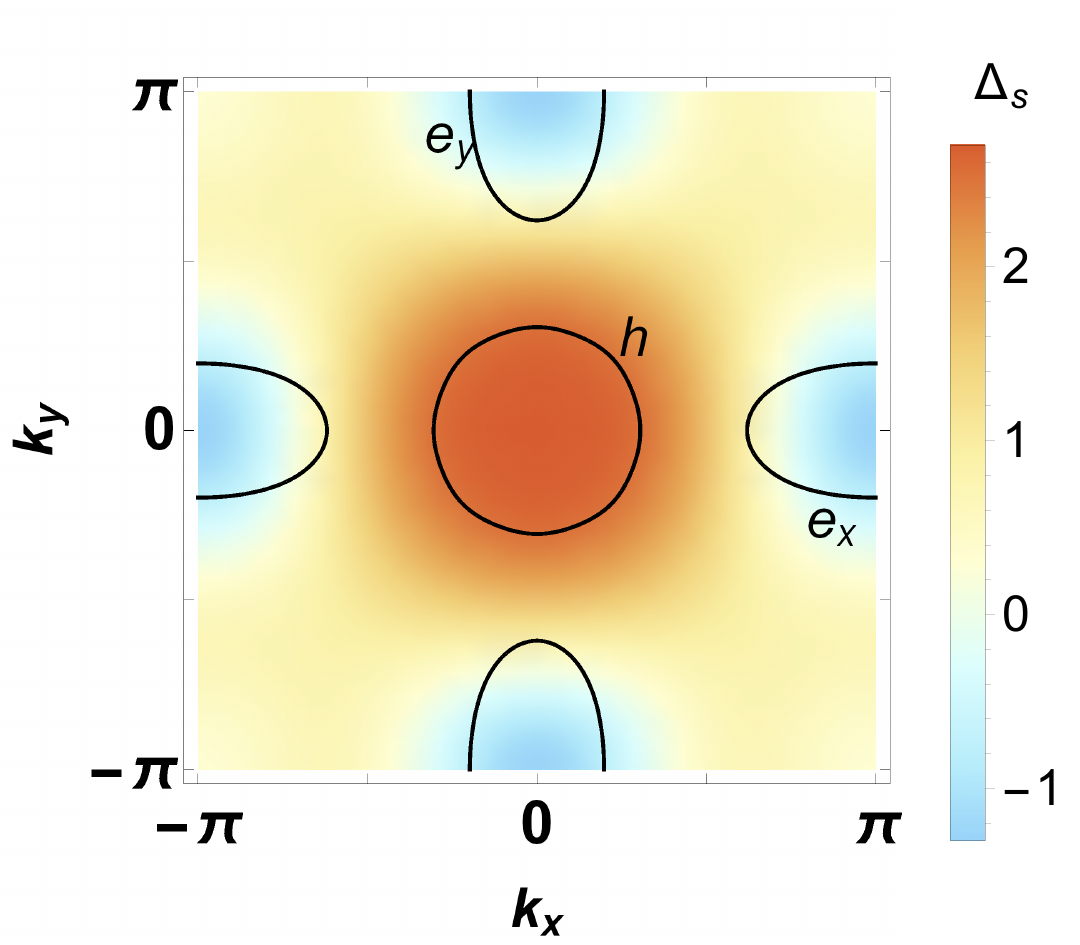}\vspace{-0.7cm}
\includegraphics[width=0.6\columnwidth]{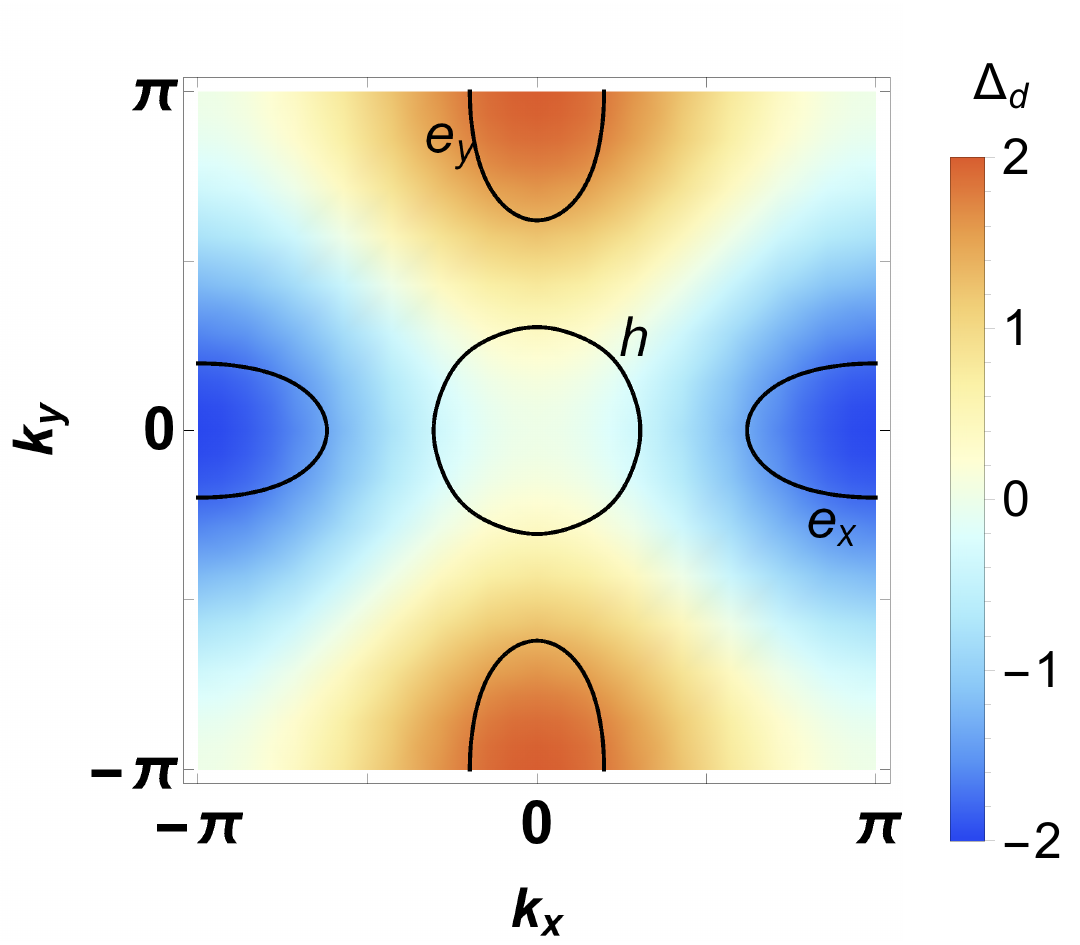} 
\par\end{centering}

\protect\protect\caption{\textbf{Schematic representation of the nodal $s^{+-}$ and $d$-wave
states. }In both panels, a density plot of the gap function is superimposed
to a representative Fermi surface consisting of a hole pocket (h)
at the center and an electron pocket (e) at the borders of the Brillouin
zone. In the nodal $s^{+-}$ states (upper panel), the nodes are not
enforced by symmetry (here they are located at the electron pockets).
In the $d$-wave state (lower panel), the nodes are enforced by symmetry
to be on the diagonals of the Brillouin zone, and therefore can only
cross the hole pockets. \label{fig4a}}
\end{figure}

To understand the implications of the fitting to a constant and an
angle-dependent gap for finite pressures, we analyze the two different
scenarios in which nodes can emerge: a nodal $s^{+-}$ state (with
gap functions of different signs in the hole and in the electron pockets)
and a $d$-wave state. In the former, the position of the nodes are
accidental, i.e. not enforced by symmetry, while in the latter the
nodes are enforced by symmetry to be on the Brillouin zone diagonals.
Schematic representations of both scenarios are shown in Fig. \ref{fig4a},
where a density plot of the gap functions is superimposed to the typical
Fermi surface of the iron pnictides, consisting of one or more hole
pockets at the center of the Brillouin zone, and electron pockets
at the border of the Brillouin zone. In this figure, we set the accidental
nodes of the $s^{+-}$ state to be on the electron pockets, as observed
by ARPES in the related compound BaFe$_{2}$(As$_{1-x}$P$_{x}$)$_{2}$
\cite{ZhangN}. Note that in the $d$-wave state, while nodes appear
in the hole pockets, the electron pockets have nearly uniform gaps.
Thus, the fact that the fitting to the $\alpha$-model gives a constant
and an angle-dependent gap is consistent with a $d$-wave state.

To contrast the scenarios of a nodal $s^{+-}$ gap and a $d$-wave
gap, we consider a microscopic model that goes beyond the simplifications
of independent gap functions of the $\alpha$-model discussed above.
In this microscopic model, the fully coupled non-linear gap equations
are solved for 
a hole pocket $h$ and two electron pockets $e_{1,2}$, and the penetration
depth is calculated at all temperatures. The free parameters are then
the density of states of the pockets, the amplitude of the pairing
interaction, and the gap functions themselves (details in the supplementary
material). For simplicity, the anisotropies of the electron pockets
are neglected, the Fermi velocities of the pockets are assumed to
be nearly the same, and the gaps are expanded in their leading harmonics.
Thus, for the nodal $s^{+-}$ state we have $\Delta_{h}=\Delta_{h,0}$
and $\Delta_{e_{i}}=\Delta_{e,0}\left(r\pm\cos2\varphi_{e}\right)$,
whereas for the $d$-wave state it follows that $\Delta_{h}=\Delta_{h,0}\cos2\varphi_{h}$
and $\Delta_{e_{i}}=\pm\Delta_{e,0}$. Note the difference in the
position of the nodes in each case: while for the $d$-wave case they
are always at $\varphi_{h}=\pm\pi/4$, for the nodal $s^{+-}$ the
nodes exist only when $r<1$ at arbitrary positions $\varphi_{e}=\pm\frac{1}{2}\arccos r$.
The results of the fittings for the pressures $p$ = 1.57 GPa and
$p=$2.25 GPa imposing a nodal $s^{+-}$ state are shown in Figs.~3c
and f. Remarkably, we find in both cases that the best fit gives 
 $r\rightarrow0$. This extreme case is, within our model, indistinguishable
from the fitting to the $d$-wave state, since in both cases the nodes
are at $\varphi=\pm\pi/4$ (albeit in different Fermi pockets). 
We  note  that  from  the  fits  one cannot  completely  rule  out  the  possibility  of  small  but
non-vanishing values of $r$. Therefore, at least within our model,  a  nodal
$s^{+-}$ state is compatible with the data  only if the accidental nodes are fine-tuned
to lie either at or very close to the diagonals of the electron pockets for a broad pressure range.
Since the position of the accidental nodes is expected to be sensitive to the topology of the Fermi surface, and consequently to pressure, it seems more plausible that the gap state is $d$-wave, since in that case 
the position of the gaps is enforced by symmetry to be along the diagonals of the hole pockets
regardless of the value of the pressure.

\begin{figure}[t!]
\centering \includegraphics[width=0.7\linewidth]{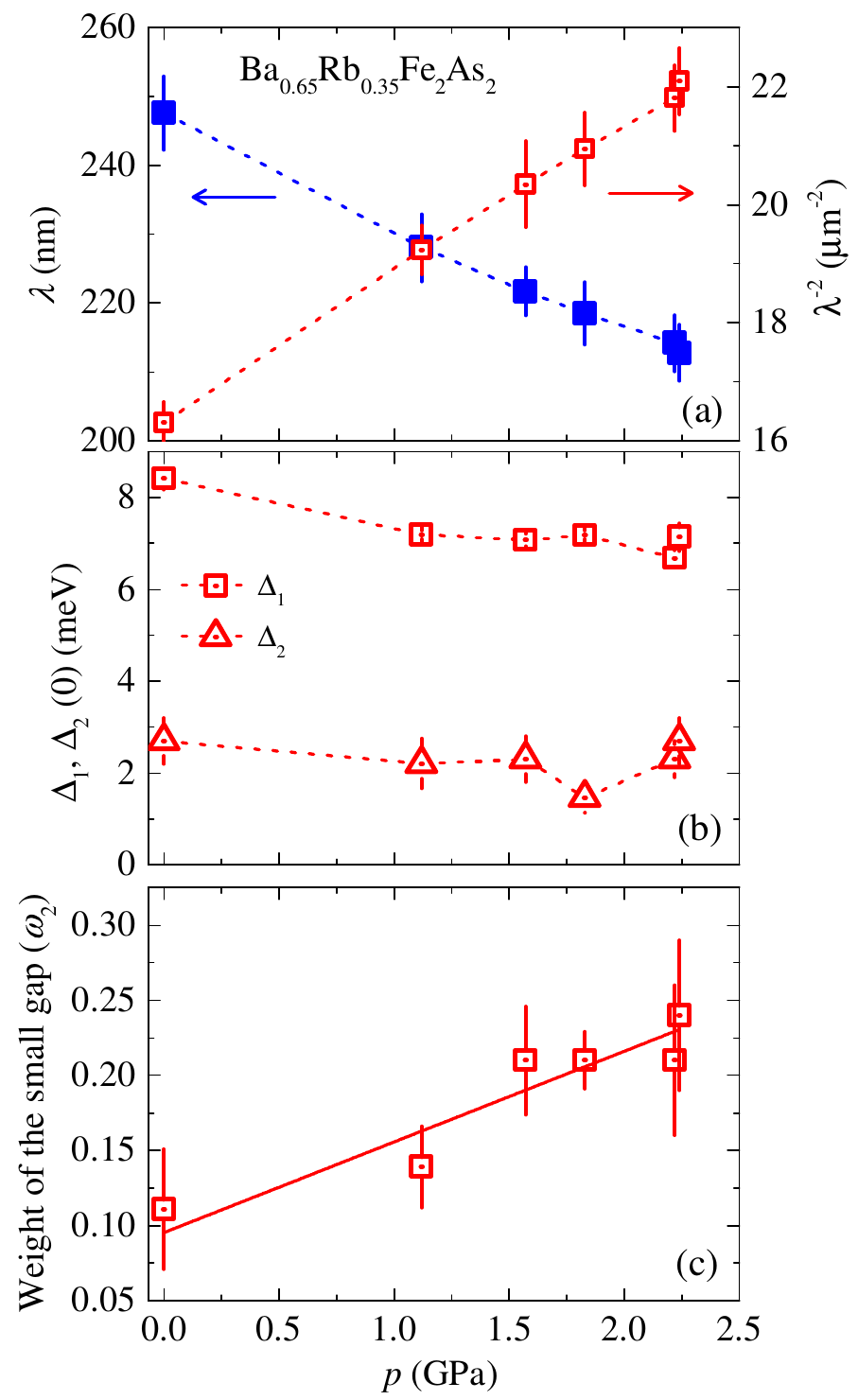} \vspace{-0.8cm}
 \protect\protect\caption{ (Color online) \textbf{Pressure dependence of various quantities
of Ba$_{0.65}$Rb$_{0.35}$Fe$_{2}$As$_{2}$.} (a) The magnetic penetration
depth ${\lambda}(0)$ and ${\lambda^{-2}}(0)$. (b) The zero-tempearture
gap values ${\Delta}_{1,2}$(0). 
The relative weight ${\omega}_{2}$ of the small gap to the superfluid
density. The dashed lines are guides to the eyes, and the solid lines
represent linear fits to the data.}

\label{fig5} 
\end{figure}


The pressure dependence of all the parameters extracted from the data
analysis within the ${\alpha}$ model are plotted in Figs.~5(a-c).
From Fig.~5a a substantial decrease of ${\lambda}\left(0\right)$
with pressure is evident. At the maximum applied pressure of $p$
= 2.25 GPa the reduction of ${\lambda}\left(0\right)$ is approximately
15 ${\%}$ compared to the value at $p$ = 0 GPa. Both ${\Delta}_{1}$
and ${\Delta}_{2}$ show a small reduction upon increasing the pressure
from $p$ = 0 to 1.17 GPa, while above $p$ = 1.17 GPa the gaps values
stay constant. On the other hand, the relative contribution ${\omega}_{2}$
of the small gap to the superfluid density increases by approximately
factor of 2 for the maximum applied pressure of $p$ = 2.25 GPa (see
Fig.~5c), indicating a spectral weight shift to the smaller gap.
The parameters extracted from the microscopic model are discussed
in the supplementary material. 

\section{DISCUSSION}

The most essential finding of our paper is the observation that pressure
promotes a nodal SC gap in Ba$_{0.65}$Rb$_{0.35}$Fe$_{2}$As$_{2}$.
This conclusion is robust and model-independent, as it relies on the
qualitative change in the low-temperature behavior of $\Delta\lambda^{-2}$
from exponential to linear in $T$ upon applied pressure. To our knowledge
this is the first direct experimental demonstration of a pressure
induced change in the superconducting gap structure in a 
Fe-HTS's. Two possible gap structures could be realized at finite pressures:
a nodal $s^{+-}$ state and a $d$-wave state. In the first case,
the change from nodeless $s^{+-}$ to nodal $s^{+-}$ is a crossover
rather than a phase transition \cite{Fernandes11,Stanev11}, whereas
in the latter it is an actual phase transition that could harbor exotic
pairing states, such as $s+id$ \cite{Thomale11,Khodas12,Fernandes13}
or $s+d$ \cite{Fernandes2_2013}.

Additional results provide important clues of how pressure may induce
either an $s^{+-}$ or a $d$-wave state. In the closely related optimally-doped
compound Ba$_{0.6}$K$_{0.4}$Fe$_{2}$As$_{2}$, Raman spectroscopy
\cite{Bohm}, as well as theoretical calculations \cite{Thomale11,Maiti11},
reveal a sub-dominant $d$-wave state close in energy to the dominant
$s^{+-}$ state. Pressure may affect this intricate balance, and tip
the balance in favor of the $d$-wave state. On the other hand, theoretical
calculations have shown that the pnictogen height is an important
factor in determining the structure of the $s^{+-}$ SC order parameter
\cite{Kuroki}. A systematic comparison of the quasiparticle excitations
in the 1111, 122, and 111 families of Fe-HTS's showed that the nodal
$s^{+-}$ state is favored when the pnictogen height decreases below
a threshold value of ${\simeq}$ 1.33 {\AA } \cite{Hashimoto-pnictogen}.
Hydrostatic pressure may indeed shorten the pnictogen height and consequently
modify the $s^{+-}$ gap structure from nodeless to nodal. Although
our fitting of the penetration depth data to both a microscopic model
and an effective $\alpha$-model suggest that the $d$-wave state
is more likely to be realized than the nodal $s^{+-}$ state, further
quantitative calculations of the pressure effect are desirable to
completely discard a nodal $s^{+-}$ state.

Besides the appearance of nodes with pressure, another interesting
observation is the reduction of ${\lambda}(0)$ under pressure, despite
the fact that $T_{c}$ remains nearly unchanged. Interestingly, in
the compound BaFe$_{2}$As$_{2-x}$P$_{x}$, a sharp enhancement of
${\lambda}(0)$ is observed as optimal doping is approached from the
overdoped side \cite{Hashimoto-Science}, which has been interpreted
in terms of a putative quantum critical point (QCP) inside the SC
dome \cite{Walsmley,Levchenko13,Sachdev}. In Ba$_{0.65}$Rb$_{0.35}$Fe$_{2}$As$_{2}$,
if such a putative QCP is also present, pressure is likely to move
the system away from the putative QCP, which, according to the results
of BaFe$_{2}$As$_{2-x}$P$_{x}$, would explain the observed suppression
of the penetration depth at $T=0$. This scenario does not explain
why $T_{{\rm c}}$ stays nearly constant under pressure, but this
could be due to the intrinsic flatness of $T_{c}$ around optimal
doping in Ba$_{1-x}$Rb$_{x}$Fe$_{2}$As$_{2}$. Note that a similar
behavior for $\lambda\left(0\right)$ and $T_{c}$ with pressure has
been recently observed in LaFeAsO$_{1-x}$F$_{x}$ \cite{Giacomo},
but interpreted in terms of the interplay between impurity scattering
and pressure. To distinguish unambiguously between these two scenarios,
pressure-dependent studies of the quasiparticle mass in Ba$_{0.65}$Rb$_{0.35}$Fe$_{2}$As$_{2}$
are desirable, in order to probe whether a putative QCP is present
or not in this compound.

\section{CONCLUSIONS}

In conclusion, the zero-temperature magnetic penetration depth ${\lambda}\left(0\right)$
and the temperature dependence of ${\lambda^{-2}}$ were studied in
optimally doped Ba$_{0.65}$Rb$_{0.35}$Fe$_{2}$As$_{2}$ by means
of ${\mu}$SR experiments as a function of pressure up to p ${\simeq}$
2.25 GPa. The SC transition temperature stays nearly constant under
pressure, whereas a strong reduction of ${\lambda}\left(0\right)$
is observed, possibly related to the presence of a putative quantum
critical point. Our main result is the demonstration that in the investigated
Fe-based superconductor a nodal SC gap is promoted by hydrostatic
pressure. Model calculations favor a $d$-wave over a nodal
$s^{+-}$-wave pairing as the origin for the nodal gap. The present
results offer important benchmarks for the elucidation of the complex
microscopic mechanism responsible for the observed non-universaltiy
of the SC gap structure and of high-temperature superconductivity
in the Fe-HTS's in general.

\section{METHODS}

\textbf{Sample}: Polycrystalline samples of Ba$_{0.65}$Rb$_{0.35}$Fe$_{2}$As$_{2}$
were prepared in evacuated quartz ampoules by a solid state reaction
method. Fe$_{2}$As, BaAs, and RbAs were obtained by reacting high
purity As (99.999 $\%$), Fe (99.9$\%$), Ba (99.9$\%$), and Rb (99.95$\%$)
at 800 $^{\circ}$C, 650 $^{\circ}$C and 500~$^{\circ}$C, respectively.
Using stoichiometric amounts of BaAs or RbAs and Fe$_{2}$As, the
terminal compounds BaFe$_{2}$As$_{2}$ and RbFe$_{2}$As$_{2}$ were
synthesized at 950 $^{\circ}$C and 650 $^{\circ}$C, respectively.
Finally, samples of Ba$_{1-x}$Rb$_{x}$Fe$_{2}$As$_{2}$ with $x$
= 0.35 were prepared from appropriate amounts of single-phase BaFe$_{2}$As$_{2}$
and RbFe$_{2}$As$_{2}$. The components were mixed, pressed into
pellets, placed into alumina crucibles, and annealed for 100 hours under vacuum
at 650 $^{\circ}$C with one intermittent grinding. Powder X-ray diffraction
analysis revealed that the synthesized samples are single phase materials.

\textbf{Pressure cell}: Pressures up to 2.4 GPa were generated in
a double wall piston-cylinder type of cell made of MP35N material,
especially designed to perform ${\mu}$SR experiments under pressure
\cite{Maisuradze,Andreica}. As a pressure transmitting medium Daphne
oil was used. The pressure was measured by tracking the SC transition
of a very small indium plate by AC susceptibility. The filling factor
of the pressure cell was maximized. The fraction of the muons stopping
in the sample was approximately 40 ${\%}$.

\textbf{${\mu}$SR experiment}: The measurements were performed using
high-pressure ${\mu}$SR technique, where an intense high-energy ($p_{\mu}$
= 100 MeV/c) beam of muons is implanted in the sample through the
pressure cell. In a ${\mu}$SR experiment nearly 100 ${\%}$ spin-polarized
muons ${\mu}$$^{+}$ are implanted into the sample one at a time.
The positively charged ${\mu}$$^{+}$ thermalize at interstitial
lattice sites, where they act as magnetic microprobes. In a magnetic
material the muon spin precesses in the local field $B_{{\rm \mu}}$
at the muon site with the Larmor frequency ${\nu}_{{\rm \mu}}$ =
$\gamma_{{\rm \mu}}$/(2${\pi})$$B_{{\rm \mu}}$ (muon gyromagnetic
ratio $\gamma_{{\rm \mu}}$/(2${\pi}$) = 135.5 MHz T$^{-1}$). By
means of $\mu$SR important length scale of superconductor can be
measured, namely the magnetic penetration depth $\lambda$. When a
type II superconductor is cooled below $T_{{\rm c}}$ in an applied
magnetic field ranging between the lower ($H_{c1}$) and the upper
($H_{c2}$) critical field, a vortex lattice is formed which in general
is incommensurate with the crystal lattice, and the vortex cores will
be separated by much larger distances than those of the unit cell.
Because the implanted muons stop at given crystallographic sites,
they will randomly probe the field distribution of the vortex lattice.
Such measurements need to be performed in a field applied perpendicular
to the initial muon spin polarization (so called TF configuration).

\textbf{Analysis of TF-${\mu}$SR data}: Our zero-field ${\mu}$SR
experiments (see supplemental material) reveal a magnetic fraction
of about 10 ${\%}$ in the sample, caused by the presence of diluted
Fe moments as discussed in previous ${\mu}$SR studies. The signal
from the magnetically ordered parts vanishes within the first 0.2
${\mu}$s. Thus, the fits of TF data were restricted to times $t$
${>}$ 0.2 ${\mu}$s for all temperatures.

The TF ${\mu}$SR data were analyzed by using the following functional
form:\cite{Bastian} 
\begin{equation}
\begin{aligned}P(t)=A_{s}\exp\Big[-\frac{(\sigma_{sc}^{2}+\sigma_{nm}^{2})t^{2}}{2}\Big]\cos(\gamma_{\mu}B_{int,s}t+\varphi)\\
+A_{pc}\exp\Big[-\frac{\sigma_{pc}^{2}t^{2}}{2}\Big]\cos(\gamma_{\mu}B_{int,pc}t+\varphi),
\end{aligned}
\end{equation}
and $A_{{\rm pc}}$ denote the initial assymmetries of the sample
and the pressure cell, respectively. $\gamma/(2{\pi})\simeq135.5$~MHz/T
is the muon gyromagnetic ratio, ${\varphi}$ is the initial phase
of the muon-spin ensemble, and $B_{{\rm int}}$ represents the internal
magnetic field at the muon site. The relaxation rates ${\sigma}_{{\rm sc}}$
and ${\sigma}_{{\rm nm}}$ characterize the damping due to the formation
of the vortex lattice in the SC state and of the nuclear magnetic
dipolar contribution, respectively. In the analysis ${\sigma}_{{\rm nm}}$
was assumed to be constant over the entire temperature range and was
fixed to the value obtained above $T_{{\rm c}}$, where only nuclear
magnetic moments contribute to the muon relaxation rate ${\sigma}$.
The Gaussian relaxation rate ${\sigma}_{{\rm pc}}$ reflects the depolarization
due to the nuclear magnetism of the pressure cell. It can be seen
from the FT's shown in Figs.~1c and d that the width of the pressure
cell signal increases below $T_{c}$. As shown previously \cite{Maisuradze-PC},
this is due to the influence of the diamagnetic moment of the SC sample
on the pressure cell, leading to a temperature dependent ${\sigma}_{{\rm pc}}$
below $T_{c}$. In order to consider this influence, we assume a linear
coupling between ${\sigma}_{{\rm pc}}$ and the field shift of the
internal magnetic field in the SC state: ${\sigma}_{{\rm pc}}$($T$)
= ${\sigma}_{{\rm pc}}$($T$ {\textgreater} $T_{{\rm c}}$) +
$C(T)$(${\mu}_{{\rm 0}}$$H_{{\rm int,NS}}$ - ${\mu}_{{\rm 0}}$$H_{{\rm int,SC}}$),
where ${\sigma}_{{\rm pc}}$($T$ {\textgreater} $T_{{\rm c}}$)
= 0.35 ${\mu}$$s^{-1}$ is the temperature independent Gaussian relaxation
rate. ${\mu}_{{\rm 0}}$$H_{{\rm int,NS}}$ and ${\mu}_{{\rm 0}}$$H_{{\rm int,SC}}$
are the internal magnetic fields measured in the normal and in the
SC state, respectively. As indicated by the solid lines in Figs.~1a-d,
the ${\mu}$SR data are well described by Eq.~(1). The solid lines
in panels c and d are the FTs of the fitted curves shown in Figs.~1a
and b. The model used describes the data rather well.


\textbf{\section{Acknowledgments}}

Experimental work was performed at the Swiss Muon Source (S${\mu}$S)
Paul Scherrer Insitute, Villigen, Switzerland. 
Z.G. acknowledge the support by the Swiss National Science Foundation.
R.M.F and J.K. were support by the U.S. Department of Energy, Office of Science, Basic Energy
Sciences, under award number DE-SC0012336. 
A.S. acknowledges support from the SCOPES grant No. IZ74Z0-137322.
G.P. is supported by the Humboldt Research Fellowship for Postdoctoral Researchers. 

\textbf{\section{Contributions}}
Project planning:  Z.G.; Sample growth: Z.B.;
${\mu}$SR experiments:  Z.G.; R.K.; A.A.; H.L.; P.K.B.; E.M; A.S.; G.P.; H.K., and F.V.R.;
Magnetization experiment: Z.G., and F.V.R.;
${\mu}$SR data analysis: Z.G.; Analysis of the penetration depth data with ${\alpha}$-model: Z.G.;
Analysis of the penetration depth data with the microscopic model: J.K., and R.M.F.; 
Data interpretation: Z.G.; R.M.F., and R.K.;
Draft writing: Z.G., with contributions and/or comments from all authors.

\newpage
{SUPPLEMENTAL MATERIAL} 


\subsection{Experimental Details}

\subsubsection{Sample characterization}

\begin{figure}[b!]
\includegraphics[width=0.6\linewidth]{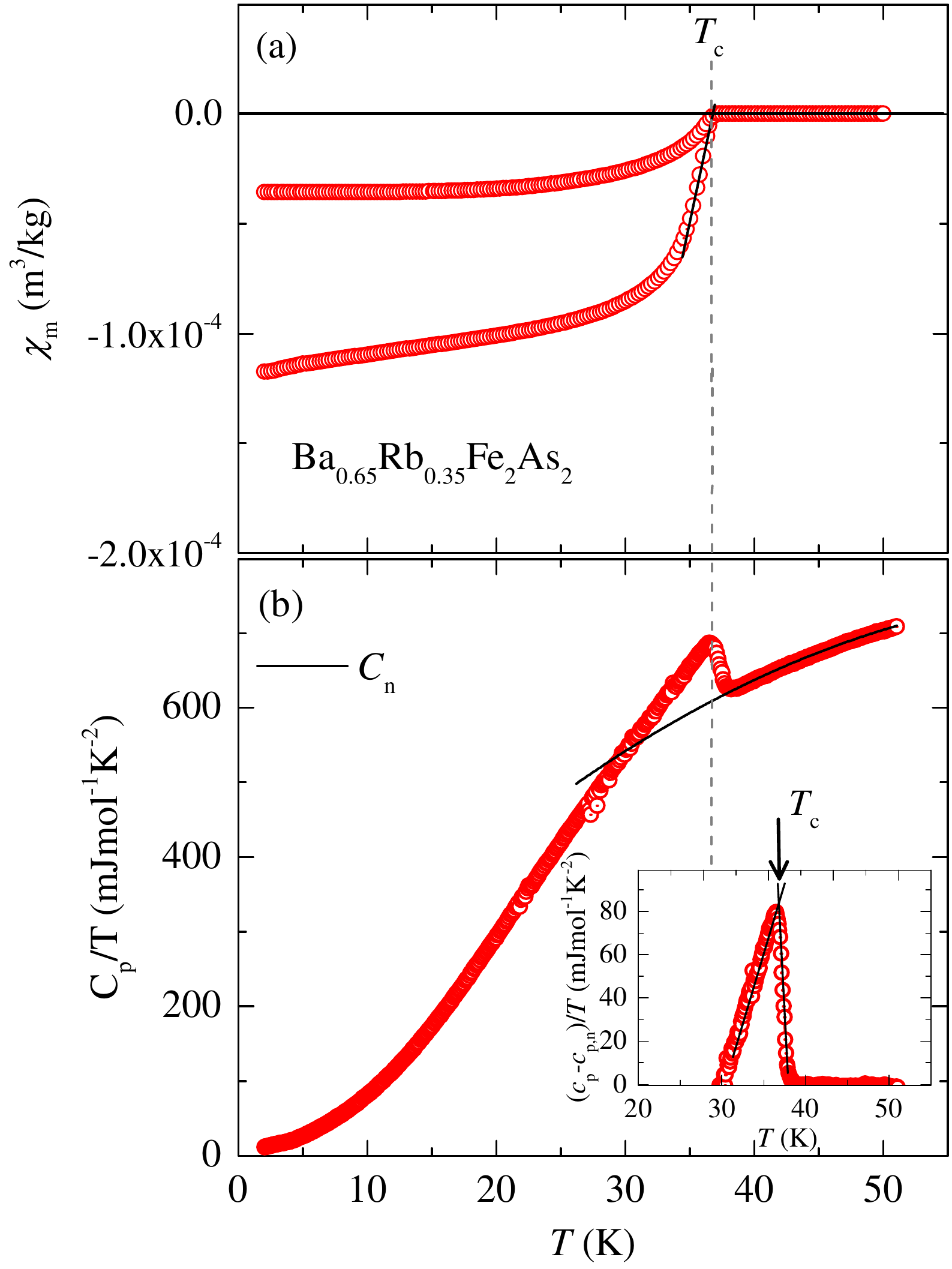}
\protect\caption{ (Color online) (a) Temperature dependence of the zero-field cooled
(ZFC )and field-cooled (FC) susceptibility ${\chi_{{\rm m}}}$ obtained
in an applied magnetic field of $\mu_{0}$$H$ = 10 mT for optimally
doped Ba$_{0.65}$Rb$_{0.35}$Fe$_{2}$As$_{2}$. (b) Specific heat
$C_{{\rm p}}$/$T$ as a function of temperature of Ba$_{0.65}$Rb$_{0.35}$Fe$_{2}$As$_{2}$.
The arrow denotes the superconducting transition temperature $T_{c}$.}

\label{fig1} 
\end{figure}

\begin{figure}[t!]
\includegraphics[width=0.8\linewidth]{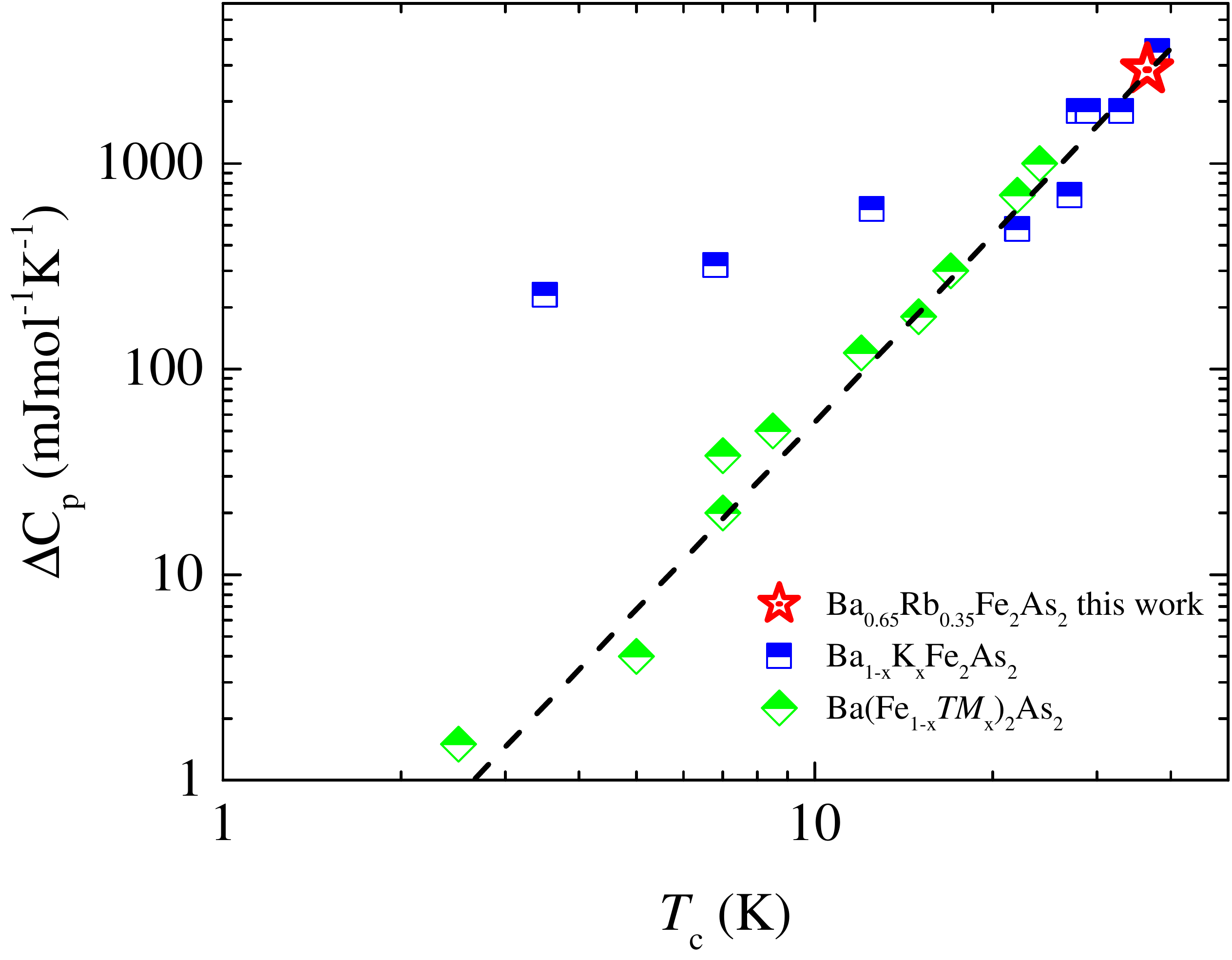} \protect\caption{ (Color online) (a) Specific heat jump $\Delta$$C_{{\rm p}}$ at
the superconducting transition vs $T_{{\rm c}}$ for Ba$_{0.65}$Rb$_{0.35}$Fe$_{2}$As$_{2}$,
plotted together with literature data for various FeAs-based superconductors.
The line corresponds to $\Delta$$C_{{\rm p}}$ ${\propto}$ $T^{3}$
(after {[}1{]}).}

\label{fig1} 
\end{figure}

The temperature dependence of the zero field-cooled
(ZFC) and field-cooled (FC) diamagnetic susceptibility of Ba$_{0.65}$Rb$_{0.35}$Fe$_{2}$As$_{2}$
measured in a magnetic field of $\mu_{{\rm 0}}$$H$ = 1 mT is shown
in Fig. 6(a). From the diamagnetic response the SC transition temperature
$T_{{\rm c}}$ is determined from the intercept of the linearly extrapolated
zero-field cooled (ZFC) susceptibility curve with $\chi_{m}$ = 0
line, and it is found to be $T_{{\rm c}}$ = 36.8(5) K. The temperature-dependent
heat capacity data for this sample plotted as $C_{{\rm p}}$/$T$
vs $T$ is shown in Fig.~6(b). The jump associated with the SC transitions
is clearly seen. Here the anomaly at the transition has been isolated
from the phonon dominated background by subtracting a second order
polynomial $C_{{\rm p,n}}$ fitted above $T_{{\rm c}}$ and extrapolated
to lower temperature. The quantity ${\Delta}C_{{\rm p}}$/$T$ with
${\Delta}C_{{\rm p}}$ = ($C_{{\rm p}}$ - $C_{{\rm p,n}}$) is presented
as a function of temperature in the inset of Fig.~6(b). Although
there may be some uncertainty in using this procedure over an extended
temperature range, the lack of appreciable thermal SC fluctuations,
as evidenced by the mean-field-like form of the anomaly, means that
there is very little uncertainty in the size of ${\Delta}C_{{\rm p}}$.
Bud'ko et.~al. \cite{Budko} found that in many '122' Fe-based superconductors
the specific heat jump ${\Delta}C_{{\rm p}}$ at $T_{{\rm c}}$ follows
the empirical trend, the so-called BNC scalling ${\Delta}C_{{\rm p}}$
${\propto}$ $T^{3}$. This has been interpreted as either originating
from quantum critically or from strong impurity pair breaking. A violation
of the BNC scaling was observed for Ba$_{1-x}$K$_{x}$Fe$_{2}$As$_{2}$
for $x$ ${\textgreater}$ 0.7 \cite{Budko} and in addition a change
of the SC gap symmetry was observed. The specific heat jump data for
Ba$_{0.65}$Rb$_{0.35}$Fe$_{2}$As$_{2}$ obtained in this work is
added in Fig.~7 to the BNC plot taken from Ref.~\cite{Budko}. Our
data point lies perfectly on the BNC line.


\subsubsection{Muon spin rotation experiments}

Zero-field (ZF) and transverse-field (TF) ${\mu}$SR experiments at
ambient and under various applied pressures were performed at the
$\mu$E1 beamline of the Paul Scherrer Institute (PSI), Switzerland,
using the dedicated GPD spectrometer. A gas-flow $^{4}$He (base temperature
${\sim}$ 4 K) and a VARIOX cryostat (base temperature ${\sim}$ 1.3
K) were used. High energy muons ($p_{\hat{I}\frac14}$ = 100 MeV/c)
were implanted in the sample. Forward and backward positron detectors
with respect to the initial muon spin polarization were used for the
measurements of the ${\mu}$SR asymmetry time spectrum $A$($t$).
The typical statistics for both forward and backward detectors were
6 millions. All ZF and TF ${\mu}$SR experiments were performed by
stabilizing the temperature in prior to recording the ${\mu}$SR-time
spectra. Note that a precise calibration of the GPD results was carried
out at the ${\pi}$M3 beamline using the low background general purpose
instrument (GPS). The ${\mu}$SR time spectra were analyzed using
the free software package MUSRFIT \cite{Suter}.

 In a ${\mu}$SR experiment nearly 100 ${\%}$ spin-polarized muons ${\mu}$$^{+}$ are implanted into the sample one at a time. The positively charged ${\mu}$$^{+}$ thermalize at interstitial lattice sites, where they act as magnetic microprobes. In a magnetic material the muon spin precesses either in the local or applied magnetic field $B_{\rm \mu}$ at the
penetration depth $\lambda$ and the coherence length $\xi$. If a
type II superconductor is cooled below $T_{{\rm c}}$ in an applied
magnetic field ranged between the lower ($H_{c1}$) and the upper
($H_{c2}$) critical fields, a vortex lattice is formed which in general
is incommensurate with the crystal lattice and the vortex cores will
be separated by much larger distances than those of the unit cell.
Because the implanted muons stop at given crystallographic sites,
they will randomly probe the field distribution of the vortex lattice.
Such measurements need to be performed in a field applied perpendicular
to the initial muon spin polarization (so called TF configuration).

\subsubsection{Results of the Zero-Field ${\mu}$SR experiments}

\begin{figure}[b!]
\includegraphics[width=0.8\linewidth]{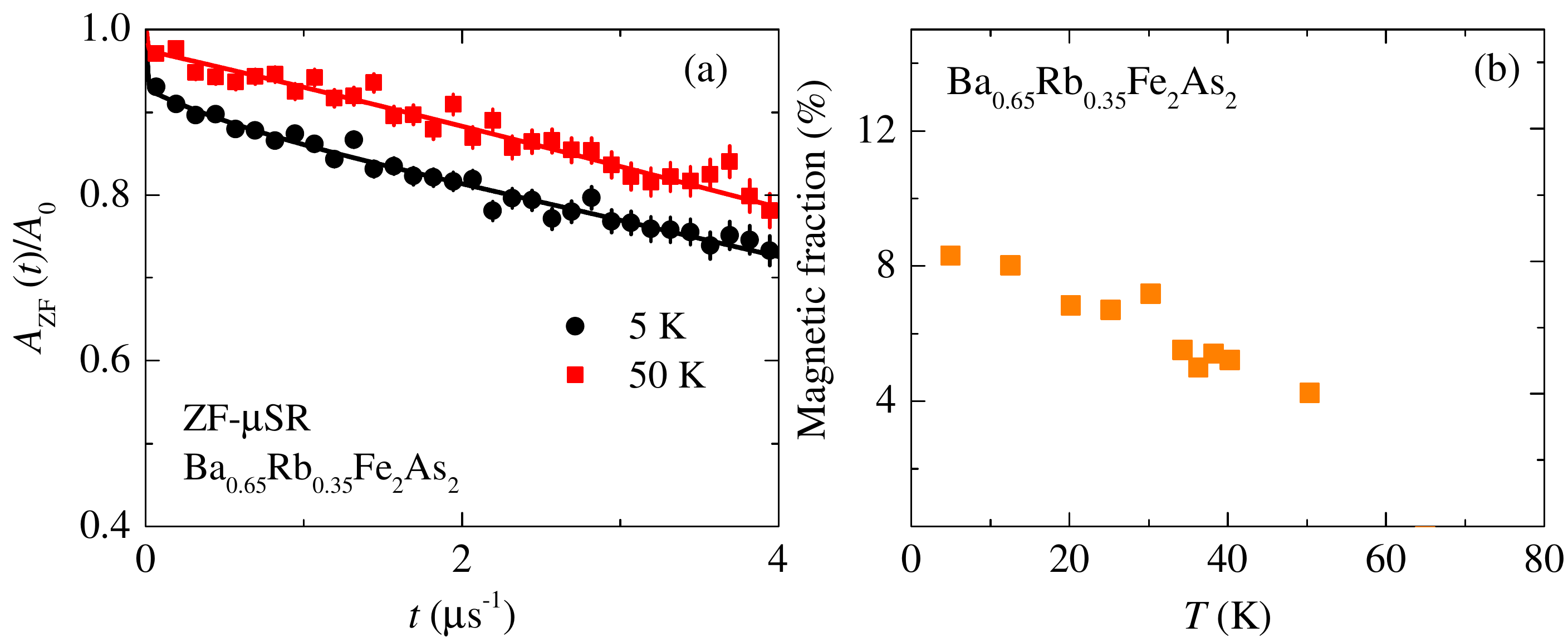} \protect\caption{(Color online) (a) The ZF-${\mu}$SR time spectra for Ba$_{0.65}$Rb$_{0.35}$Fe$_{2}$As$_{2}$
recorded above and below $T_{{\rm c}}$. The solid line represent
the fits to the data by means of Eq.~(1). (b) Temperature dependence
of the magnetic fraction of Ba$_{0.65}$Rb$_{0.35}$Fe$_{2}$As$_{2}$,
extracted from the ZF-${\mu}$SR experiments.}

\end{figure}

\begin{figure}[t!]
\center \includegraphics[width=0.8\linewidth]{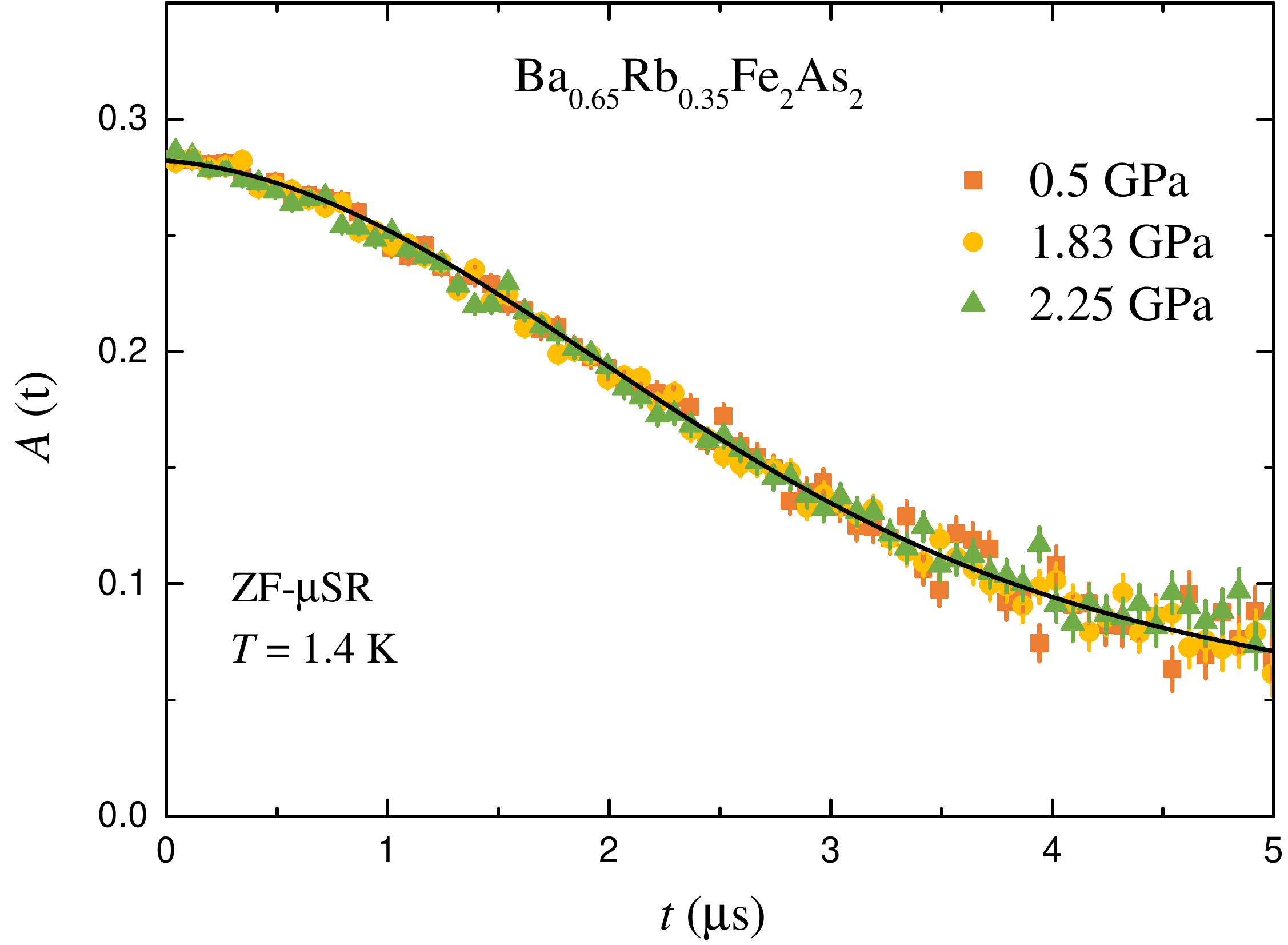}
\protect\caption{(Color online) ZF-${\mu}$SR time spectra for Ba$_{0.65}$Rb$_{0.35}$Fe$_{2}$As$_{2}$
at various applied pressures recorded at the base temperature $T$
= 1.4 K. The solid line represents the fit to the data by means of
the sum of the Eq.~(1) and a damped Kubo-Toyabe depolarization function
to account for the pressure cell signal.}

\end{figure}

It is well known that undoped BaFe$_{2}$As$_{2}$
is not superconducting at ambient pressure and undergoes a spin-density
wave (SDW) transition of the Fe-moments far above $T_{{\rm c}}$ \cite{Huang}.
The SC state can be achieved either under pressure \cite{Torikachvili,Miclea}
or by appropriate charge carrier doping of the parent compound \cite{Zhao},
leading to a suppression of the SDW state. Magnetism, if present in
the samples, must be taken into account in the TF-${\mu}$SR data
analysis. Therefore, we have carried out ZF-${\mu}$SR experiments
above and below $T_{{\rm c}}$ to search for magnetism in Ba$_{0.65}$Rb$_{0.35}$Fe$_{2}$As$_{2}$.
As an example, ZF-${\mu}$SR spectra recorded at $T$=5 K and 50 K
of Ba$_{0.65}$Rb$_{0.35}$Fe$_{2}$As$_{2}$ are shown in Fig.~8a. There is no preccesion signal, indicating that there is
no long-range magnetic order. On the other hand, we observed a significant
drop of the asymmetry, taking place within 0.2 ${\mu}$s. This is
caused by the presence of diluted Fe moments as discussed in previous
${\mu}$SR studies \cite{Khasanov-2009}. In order to quantify the magnetic
fraction, the ZF-${\mu}$SR data were analyzed by the following function:
\begin{equation}
\begin{aligned}A_{ZF}(t)={\Omega}{A_{0}}\Bigg[\frac{2}{3}e^{-\lambda_{T}t}+\frac{1}{3}e^{-\lambda_{L}t}\Bigg]\\
+(1-{\Omega}){A_{0}}\Bigg[\frac{1}{3}+\frac{2}{3}(1-{\sigma}^{2}t^{2}-{\Lambda}t)e^{{(-\frac{{\sigma^{2}}t^{2}}{2}}-{\Lambda}t)}\Bigg].
\end{aligned}
\end{equation}
the first and the second terms describe the magnetic and nonmagnetic
part of the signals, respectively. $A_{0}$ is the initial asymmetry,
${\Omega}$ is the magnetic volume fraction, and ${\lambda}_{{\rm T}}$
(${\lambda}_{{\rm L}}$) is the transverse (longitudinal) depolarization
rate of the ${\mu}$SR signal, arising from the magnetic part of the
sample. The second term describing the paramagnetic part of the sample
is the combination of a Lorentzian and a Gaussian Kubo-Toyabe depolarization
functions \cite{Kubo,Hayano}. ${\sigma}$ and ${\Lambda}$ are the
depolarization rates due to the nuclear dipole moments and randomly
oriented diluted local electronic moments, respectively. The temperature
dependence of the magnetic fraction obtained for Ba$_{0.65}$Rb$_{0.35}$Fe$_{2}$As$_{2}$
is plotted in Fig.~8b. The magnetic fraction at the base temperature
was found to be only 8 ${\%}$. Bearing in mind that the signal from
the magnetically ordered parts vanishes within the first 0.2 ${\mu}$s
in the whole temperature region, the analysis of transverse field
data was restricted to times $t$ ${\textgreater}$ 0.2 ${\mu}$s.

Figure~9 shows the ZF-${\mu}$SR time spectra for Ba$_{0.65}$Rb$_{0.35}$Fe$_{2}$As$_{2}$
at various applied pressures. The ZF relaxation rate stays nearly
unchanged between $p$ = 0 GPa and 2.25 GPa, implying that there is
no sign of pressure induced magnetism in this system.

\subsection{Microscopic model for analyzing the penetration depth data of Ba$_{0.65}$Rb$_{0.35}$Fe$_{2}$As$_{2}$}

\subsubsection{Model for $s^{+-}$ pairing}

As a minimal model that accounts for the different superconducting
states of the iron pnictides (nodeless $s^{+-}$, nodal $s^{+-}$,
and $d$-wave), we consider a two-dimensional system with three isotropic
Fermi pockets \cite{Kang14}: one hole pocket $h$ centered around
$\Gamma=(0,0)$ and two electron pockets $e_{1}$ and $e_{2}$ centered
around $M_{1}=(\pi,0)$ and $M_{2}=(0,\pi)$ (see Fig.~10). To describe
the $s^{+-}$ state, the pairing interaction between the hole pocket
$h$ and the electron pocket $e_{1}$ is assumed to be angular dependent
with the form: 
\begin{equation}
V_{he_{1}}=V_{0}(r-\cos2\phi)h_{\uparrow}^{\dagger}(\mathbf{k})h_{\downarrow}^{\dagger}(-\mathbf{k})e_{1\downarrow}(-\mathbf{p})e_{1\uparrow}(\mathbf{p})+h.c.\ ,\label{Eqn:ParingHE1}
\end{equation}
where $\phi$ is the polar angle measured relative to the center of
the electron pocket, $V_{0}$ is the interaction energy scale, and
$r$ is the relative amplitude of the angular-independent and the
angular-dependent pairing interactions. Due to the tetragonal symmetry
of the system, the pairing interaction between $h$ and $e_{2}$ is:
\begin{equation}
V_{he_{2}}=V_{0}(r+\cos2\phi)h_{\uparrow}^{\dagger}(\mathbf{k})h_{\downarrow}^{\dagger}(-\mathbf{k})e_{2\downarrow}(-\mathbf{p})e_{2\uparrow}(\mathbf{p})+h.c.\ .\label{Eqn:ParingHE2}
\end{equation}

Furthermore, to minimize the number of free parameters, we assume
that the three pockets have the same Fermi velocity $v_{f}$, while
the density of states can in principle be different $\rho_{h}/\rho_{e}=\eta$.
Within this model, we obtain an $s^{+-}$ state, where the SC gap
of the hole pocket is a constant, $\Delta_{h}$, and the gap on the
electron pockets is of the form $\Delta_{e_{1}}=\Delta_{e}(r-\cos2\phi)$
and $\Delta_{e_{2}}=\Delta_{e}(r+\cos2\phi)$. Accidental nodes appear
in the electron pockets if $r<1$. Introducing the energy cutoff $\Lambda_{c}$,
we can write down the corresponding BCS-like gap equations: 
\begin{eqnarray}
\Delta_{h} & = & -\rho_{e}V_{0}\Delta_{e}\int_{-\Lambda_{c}}^{\Lambda_{c}}\mathrm{d}\epsilon\int\frac{\mathrm{d}\phi}{2\pi}\left(\frac{(r+\cos2\phi)^{2}}{2E_{e_{1}}(\mathbf{k})}\tanh\frac{\beta E_{e_{1}}(\mathbf{k})}{2}+\frac{(r-\cos2\phi)^{2}}{2E_{e_{2}}(\mathbf{k})}\tanh\frac{\beta E_{e_{2}}(\mathbf{k})}{2}\right)\label{Eqn:BCSHole}\\
\Delta_{e} & = & -\rho_{h}V_{0}\Delta_{h}\int_{-\Lambda_{c}}^{\Lambda_{c}}\frac{\mathrm{d}\epsilon}{2E_{h}(\mathbf{k})}\tanh\frac{\beta E_{h}(\mathbf{k})}{2}\label{Eqn:BCSEle}
\end{eqnarray}
where $E_{e_{1}}(\mathbf{k})$, $E_{e_{2}}(\mathbf{k})$, and $E_{h}(\mathbf{k})$
are the quasi-particle energy dispersions: 
\[
E_{e_{1}}(\mathbf{k})=\sqrt{\epsilon_{e}^{2}+\Delta_{e}^{2}(r-\cos2\phi)^{2}}\ ,\quad E_{2}(\mathbf{k})=\sqrt{\epsilon_{e}^{2}+\Delta_{e}^{2}(r+\cos2\phi)^{2}}\ ,\quad E_{h}(\mathbf{k})=\sqrt{\epsilon_{h}^{2}+\Delta_{h}^{2}}\ .
\]

\begin{figure}[t!]
\centering \includegraphics[width=0.65\columnwidth]{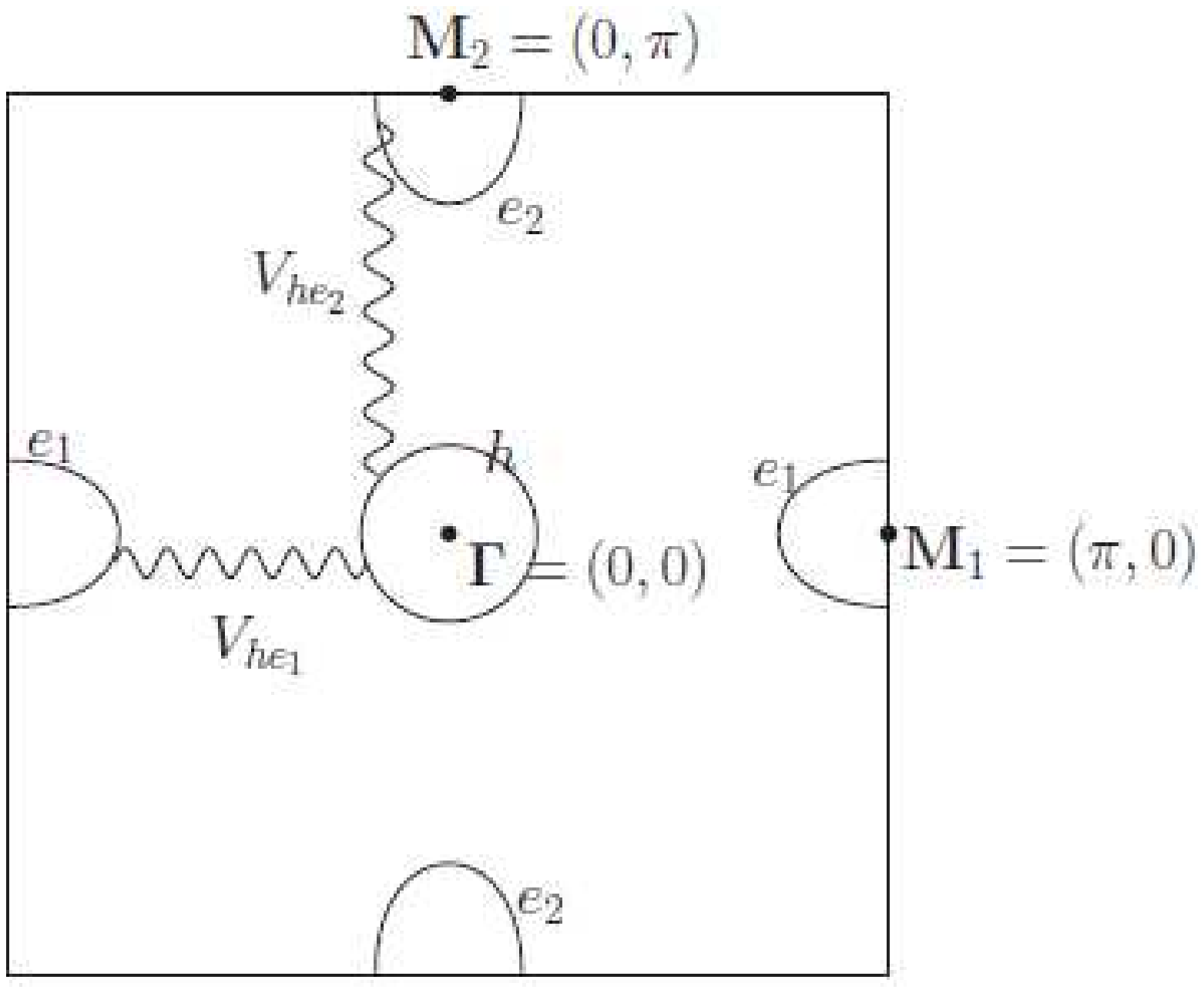}\vspace{-8cm}
\protect\caption{ (Color online) Three pocket model used in our calculations. It is
assumed that the system has one hole pocket $h$ centered around $\Gamma=(0,0)$
and two electron pockets $e_{1}$ and $e_{2}$ centered around $M_{1}=(\pi,0)$
and $M_{2}=(0,\pi)$.}

\label{fig4} 
\end{figure}

To determine $T_{c}$, we linearize the gap equations, yielding: 
\[
\left\{ \begin{array}{l}
{\displaystyle {\Delta_{h}=-\Delta_{e}\rho_{e}V_{0}(2r^{2}+1)\int_{0}^{\Lambda_{c}}\frac{\mathrm{d}\epsilon}{\epsilon}\tanh\frac{\beta_{c}\epsilon}{2}}}\\
{\displaystyle {\Delta_{e}=-\Delta_{h}\eta\rho_{e}V_{0}\int_{0}^{\Lambda_{c}}\frac{\mathrm{d}\epsilon}{\epsilon}\tanh\frac{\beta_{c}\epsilon}{2}}}
\end{array}\right.\Longrightarrow\rho_{e}V_{0}=\left[\sqrt{\eta(2r^{2}+1)}\int_{0}^{\Lambda_{c}}\frac{\mathrm{d}\epsilon}{\epsilon}\tanh\frac{\beta_{c}\epsilon}{2}\right]^{-1}
\]

To perform the fitting, we set $T_{c}$ to be fixed, and set the energy
cutoff $\Lambda_{c}=86$meV (the results do not depend significantly
on the choice of the cutoff). This provides a constraint on $\rho_{e}V_{0}$,
$\eta$, and $r$. When $T<T_{c}$, the gaps are calculated based
on the BCS Eqs. \eqref{Eqn:BCSHole} and \eqref{Eqn:BCSEle}.

The expression for the penetration depth of a single-band system is:
\[
\lambda_{\mu\mu}^{-2}(T)=\frac{4\pi}{cV}\sum_{\mathbf{k}}\left[\langle\frac{\partial^{2}\epsilon}{\partial k_{\mu}^{2}}\rangle+\left(\frac{\partial\epsilon}{\partial k_{\mu}}\right)^{2}\frac{\partial f}{\partial E_{k}}\right]\rightarrow\frac{1}{V}\sum_{\mathbf{k}}\left(\frac{\partial\epsilon}{\partial k_{\mu}}\right)^{2}\left[\frac{\partial f}{\partial E_{k}}-\frac{\partial f}{\partial\epsilon_{k}}\right]\ ,
\]
where $f$ is the Fermi distribution function, $\epsilon$ is the
energy of the non-interacting system, and $E_{k}$ is the quasi-particle
energy dispersion. Applying this formula to our three pocket model,
we obtain 
\begin{align}
\lambda^{-2}(T)\propto & \rho_{h}\frac{v_{f}^{2}}{2}\int_{-\Lambda_{c}}^{\Lambda_{c}}\mathrm{d}\epsilon\left(\frac{\partial f}{\partial E_{h}}-\frac{\partial f}{\partial\epsilon_{h}}\right)+\rho_{e}v_{f}^{2}\int_{-\Lambda_{c}}^{\Lambda_{c}}\mathrm{d}\epsilon\int\frac{\mathrm{d}\phi}{2\pi}\cos^{2}\phi\left(\frac{\partial f}{\partial E_{e_{1}}}-\frac{\partial f}{\partial\epsilon_{e}}\right)\nonumber \\
 & +\rho_{e}v_{f}^{2}\int_{-\Lambda_{c}}^{\Lambda_{c}}\mathrm{d}\epsilon\int\frac{\mathrm{d}\phi}{2\pi}\cos^{2}\phi\left(\frac{\partial f}{\partial E_{e_{2}}}-\frac{\partial f}{\partial\epsilon_{e}}\right)\nonumber \\
\lambda^{-2}(T)\propto & \rho_{e}v_{f}^{2}\left[\frac{2+\eta}{2}\Big(1-2f(\Lambda_{c})\Big)+\eta\int_{0}^{\Lambda_{c}}\mathrm{d}\epsilon\frac{\partial f}{\partial E_{h}}+2\int_{0}^{\Lambda_{c}}\mathrm{d}\epsilon\int\frac{\mathrm{d}\phi}{2\pi}\frac{\partial f}{\partial E_{e}}\right]\label{Eqn:InvPD}
\end{align}

In the fittings, we will focus on the normalized penetration depth
$\lambda^{-2}\left(T\right)/\lambda^{-2}\left(0\right)$.

\subsubsection{Model for $d$-wave pairing}

To describe the $d$-wave superconducting state within our three band
model, we consider the following form of the pairing interaction: 

\begin{align*}
V_{he_{1}} & =V_{0}(r-\cos2\theta)h_{\uparrow}^{\dagger}(\mathbf{k})h_{\downarrow}^{\dagger}(-\mathbf{k})e_{1\downarrow}(-\mathbf{p})e_{1\uparrow}(\mathbf{p})+h.c.\\
V_{he_{2}} & =V_{0}(r+\cos2\theta)h_{\uparrow}^{\dagger}(\mathbf{k})h_{\downarrow}^{\dagger}(-\mathbf{k})e_{2\downarrow}(-\mathbf{p})e_{2\uparrow}(\mathbf{p})+h.c.\ .
\end{align*}
where $\theta$ is the angle around the hole pocket. The gap functions
can then be written as: 
\[
\Delta_{e_{1}}=-\Delta_{e_{2}}=\Delta_{e}\ ,\quad\Delta_{h}(\mathbf{k})=\Delta_{h}\cos2\theta\ .
\]
resulting in the BCS-like gap equations: 
\begin{align*}
\Delta_{h} & =2\Delta_{e}\rho_{e}V_{0}\int_{-\Lambda_{c}}^{\Lambda_{c}}\frac{\mathrm{d}\epsilon}{2E_{e}}\tanh\frac{\beta E_{e}}{2}\\
\Delta_{e} & =\Delta_{h}\eta\rho_{e}V_{0}\int_{-\Lambda_{c}}^{\Lambda_{c}}\mathrm{d}\epsilon\int\frac{\mathrm{d}\theta}{2\pi}\frac{\cos^{2}2\theta}{2E_{h}}\tanh\frac{\beta E_{h}}{2}
\end{align*}

Here, $\eta=\rho_{h}/\rho_{e}$, $E_{e}=\sqrt{\epsilon_{e}^{2}+\Delta_{e}^{2}}$,
and $E_{h}=\sqrt{\epsilon^{2}+\Delta_{h}^{2}\cos^{2}2\theta}$. Repeating
the same steps as for the $s^{+-}$ case, we obtain the penetration
depth: 
\begin{equation}
\lambda^{-2}(T)\propto\rho_{e}v_{f}^{2}\left[\frac{2+\eta}{2}(1-2f(\Lambda_{c}))+\eta\int_{0}^{\Lambda_{c}}\mathrm{d}\epsilon\int\frac{\mathrm{d}\theta}{2\pi}\frac{\partial f}{\partial E_{h}}+2\int_{0}^{\Lambda_{c}}\mathrm{d}\epsilon\frac{\partial f}{\partial E_{e}}\right]
\end{equation}

Comparing the expressions for the $d$-wave case to the expressions
we derived for the $s^{+-}$ case, Eqs.~(\ref{Eqn:BCSHole}) and
(\ref{Eqn:InvPD}), we note that they can be mapped onto each other
if $r=0$. In this extreme case, changing $\eta_{d}\rightarrow4/\eta_{s}$,
$V_{0,d}\rightarrow\eta V_{0,s}/2$, and $\Delta_{h}\leftrightarrow\Delta_{e}$
leads to the same gap equations and penetration depth expression.
With these replacements, both $s$ and $d$ pairing give the same
$\lambda^{-2}(T)/\lambda^{-2}(0)$. Therefore, we conclude that the
penetration depth cannot distinguish between nodal-$s^{+-}$ and $d$-wave
if the nodal-$s^{+-}$ is the extreme case with $r=0$.

\subsubsection{Fitting Results}

We now fit the experimental data $\lambda^{-2}\left(T\right)/\lambda^{-2}\left(0\right)$
of optimally-doped Ba$_{1-x}$Rb$_{x}$Fe$_{2}$As$_{2}$ to find
the values of $\rho_{e}V_{0}$, $\eta$, and $r$ for different pressures.
Note that the value of $T_{c}$ imposes another constraint on these
three parameters, as explained above. Figs.~11a, b and c show the
fitting for the $s^{+-}$ model for $P=0$, $P=1.57$ GPa, and $P=2.25$
GPa, respectively. For the $P=0$ case, we find equal gap amplitudes
and no nodes, as seen by ARPES experiments in the related compound
Ba$_{1-x}$K$_{x}$Fe$_{2}$As$_{2}$. We see that the fitting is
not as good in the region immediately below $T_{c}$. We will discuss
this issue in more details below. For the pressurized samples, the
fitting is overall better and indicates a nodal state ($r<1$). The
value of the density of states ratio $\rho_{h}/\rho_{e}$ is little
affected by pressure (as expected, since no charge carriers are introduced),
and is consistent with the value of a nearly compensated metal. 

\begin{figure*}[b!]
\centering \includegraphics[width=1\linewidth]{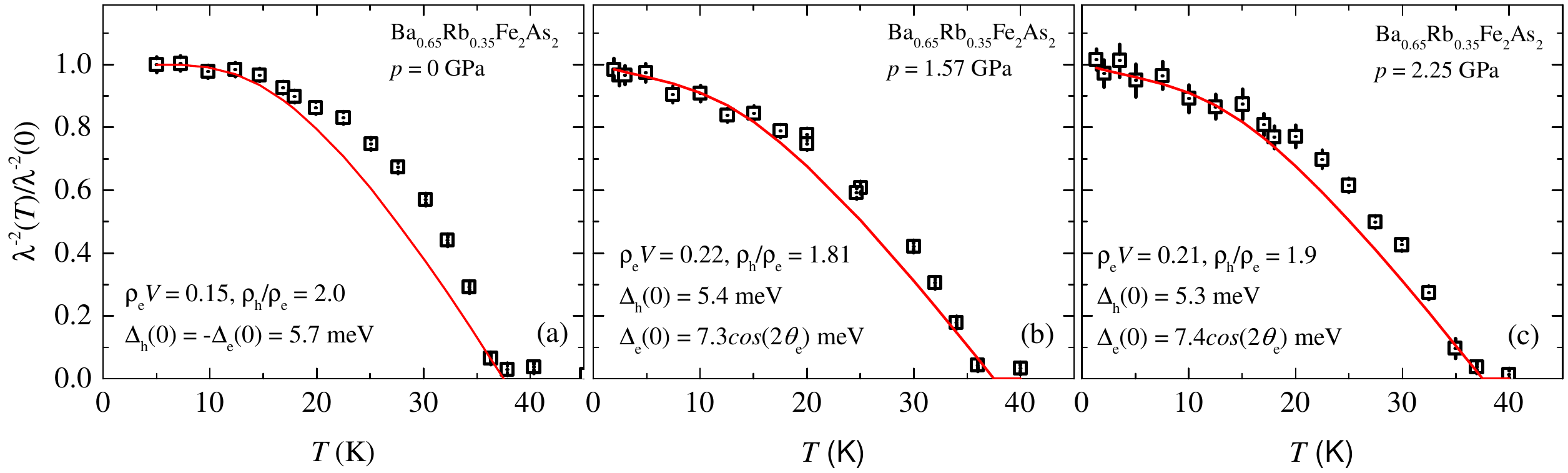}
\protect\caption{(Color online) The temperature dependence of ${\lambda}^{-2}(T)$/${\lambda}^{-2}(0)$
measured at various applied hydrostatic pressures of Ba$_{0.65}$Rb$_{0.35}$Fe$_{2}$As$_{2}$.
The square symbols are experimental data and the red curves are the
theoretical functions. (a) Fitting for the  $P=0$ data, which suggests
a nodeless state. (b) and (c) Fitting for $P=1.57$ GPa and $P=2.25$
GPa. The fitting suggests that nodes exists on the two electron pockets
at the angles $\theta_{e}=\pm\pi/4$ and $\pm3\pi/4$.}

\label{fig1} 
\end{figure*}

\begin{figure*}[t!]
\centering \includegraphics[width=1\linewidth]{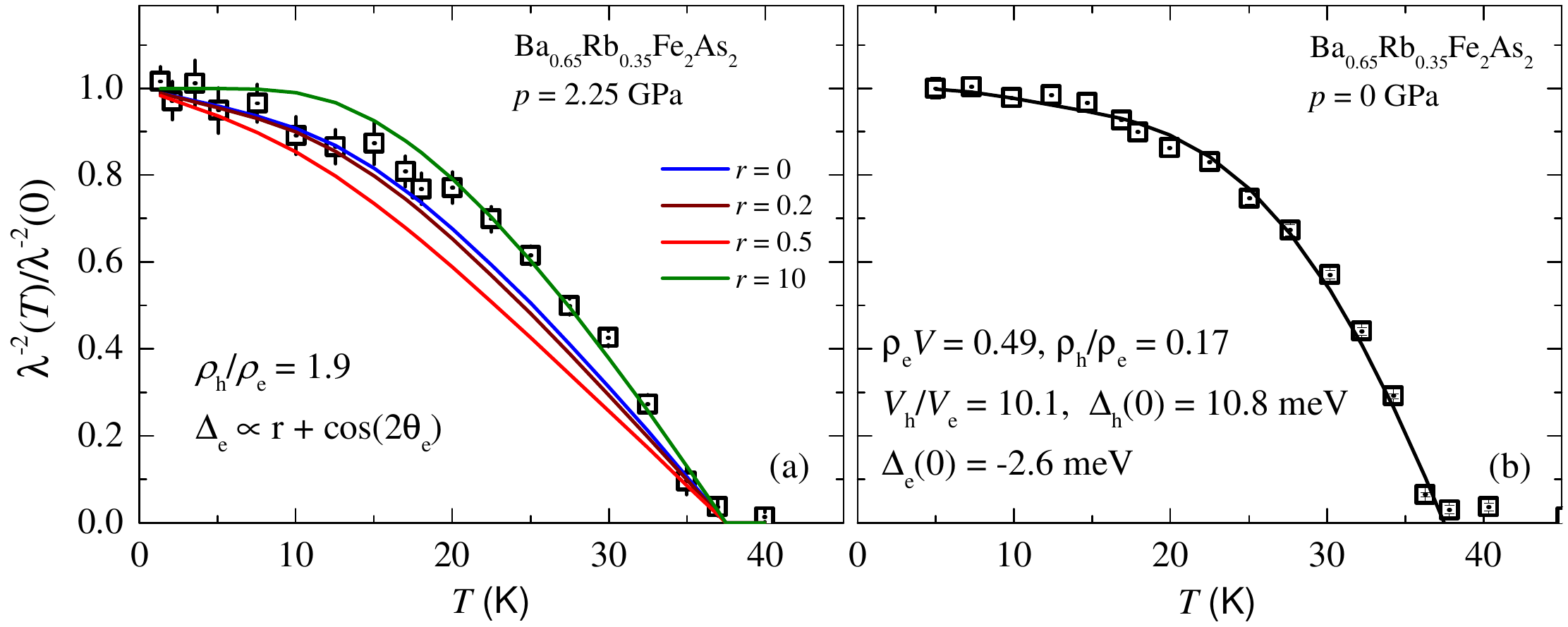}
\protect\caption{ (Color online) (a) Effect of the electron pocket gap anisotropy on
the penetration depth at $p=2.25$ GPa. The electron gap is nodal
if $r<1$, and becomes nodeless if $r>1$. The low temperature data
clearly shows that the gap is nodal, but the data near $T_{c}$ seems
to be better described by a nodeless state. (b) Fitting for the zero
pressure case with the Fermi velocity ratio $v_{h}/v_{e}$ being a
free parameter. The fitting improves with respect to Fig.~6a, but
the values of $v_{h}/v_{e}$ and $\rho_{h}/\rho_{e}$ seem to be too
large or too small.}

\label{fig1} 
\end{figure*}

Surprisingly, the best fittings for both the $P=1.57$ GPa and $P=2.25$
GPa cases give $r=0$, where the nodes on the electron pockets are
fixed at $\theta=\pm\pi/4$. This is a very special case of the accidentally
nodal $s^{+-}$ state, since by symmetry there is no reason for $r$
to vanish. To make this point more transparent, in Fig.~12a we plot
the non-zero pressure data and the theoretical urves for the penetration
depth for various values of $r$ -- keeping all the other parameters
constant. Clearly, $0<r<1$ gives worst fittings than $r=0$. What
we also found is that $r=10$ -- i.e. a nodeless superconducting state
-- describes the data better near $T_{c}$, on the expense of a very
bad fitting at low temperatures -- where the nodal behavior is evident.

As we discussed in the previous section, a nodal-$s^{+-}$ state with
$r=0$ is indistinguishable -- for fitting purposes -- from a $d$-wave
state. Since there is no symmetry reason to have $r=0$ in our simple
model, or even $r\ll1$ over a wide pressure range, we interpret this
result as an indirect indication that a $d$-wave state is more likely
to be the state of the pressurized samples. 

Finally, we comment on the difficulty of the fittings to capture the
behavior near $T_{c}$ -- particularly for the sample at ambient pressure
(see Fig.~11a). One reason could be the presence of inhomogeneities,
which would require a distribution of gaps to be taken into account,
instead of a single gap value. Another reason could be related to
our choice of fixing the Fermi velocities to be the same for both
the electron and hole pockets. To investigate this possibility, we
lift this restriction and allow $v_{h}/v_{e}$ to also be a fitting
parameter. The result is shown in Fig.~12b. Clearly, we obtain a better
fitting, but not only $\rho_{e}V_{0}$ is relatively large, but the
ratios $\rho_{h}/\rho_{e}$ and $v_{h}/v_{e}$ are very large or very
small, which is difficult to reconcile with the Fermi surface of these
materials. Most likely, additional pockets are necessary to capture
the full temperature dependence of the penetration depth. Nevertheless,
our microscopic model provides results that agree with those obtained
from the $\alpha$-model fitting, particularly in the low-temperature
regime, suggesting that a $d$-wave state is more likely to be realized
than a nodal $s^{+-}$ state.


\begin{thebibliography}{10}
\bibitem{Ding} Ding, H., Richard, P., Nakayama, K., Sugawara, T., Arakane, T.,
Sekiba, Y., Takayama, A., Souma, S., Sato, T., Takahashi, T., Wang, Z.,
Dai, X., Fang, Z., Chen, G.F., Luo, J.L. and Wang N.L. Observation of
Fermi-surface-dependent nodeless superconducting gaps in Ba$_{0.6}$K$_{0.4}$Fe$_{2}$As$_{2}$.
$Europhys.~Lett.$ \textbf{83}, 47001 (2008).

\bibitem{KhasanovN} Khasanov, R., Evtushinsky, D.V., Amato, A., 
Klauss, H.-H., Luetkens, H., Niedermayer, Ch.,  B\"{u}chner, B., Sun, G.L., Lin, C.T.,
Park, J.T., Inosov, D.S. and Hinkov, V. Two-Gap Superconductivity in
Ba$_{1-x}$K$_{x}$Fe$_{2}$As$_{2}$: A Complementary Study of the
Magnetic Penetration Depth by Muon-Spin Rotation and Angle-Resolved
Photoemission. $Phys.~Rev.~Lett.$ \textbf{102}, 187005 (2009).

\bibitem{GuguchiaN} Guguchia, Z., Shermadini, Z., Amato, A., Maisuradze, A.,
Shengelaya, A., Bukowski, Z., Luetkens, H., Khasanov, R., Karpinski, J.
and Keller, H. Muon-spin rotation measurements of the magnetic penetration
depth in the Fe-based superconductor Ba$_{1-x}$Rb$_{x}$Fe$_{2}$As$_{2}$.
$Phys.~Rev.~B$ \textbf{84}, 094513 (2011).

\bibitem{TerashimaN} Terashima, K., Sekiba, Y., Bowen, J.H., Nakayama, K.,
Kawahara, T., Sato, T., Richard, P., Xu, Y.M., Li, L.J., Cao, G.H., 
Xu, Z.A., Ding, H. and T. Takahashi, T. Fermi surface nesting induced strong
pairing in iron-based superconductors. $Proc.~Natl~Acad.~Sci.~USA$ \textbf{106},
7330-7333 (2009).

\bibitem{ZhangY} Zhang, Y., Yang, L.X., Xu, M., Ye, Z.R., Chen, F.,
He, C., Xu, H.C., Jiang, J., Xie, B.P., Ying, J.J., Wang, X.F., 
Chen, X.H., Hu, J.P., Matsunami, M., Kimura, S. and Feng, D.L. Nodeless superconducting
gap in A$_{x}$Fe$_{2}$Se$_{2}$ (A = K, Cs) revealed by angle-resolved
photoemission spectroscopy. $Nature~Mater.$ \textbf{10}, 273-277 (2011).

\bibitem{MiaoN} Miao, H., Richard, P., Tanaka, Y., Nakayama, K., Qian, T.,
Umezawa, K., Sato, T., Xu, Y.-M., Shi, Y.-B., Xu, N., Wang, X.-P., Zhang, P.,
Yang, H.-B., Xu, Z.-J., Wen, J. S., Gu, G.-D., Dai, X., Hu, J.-P., Takahashi, T. and
Ding, H. Isotropic superconducting gaps with enhanced pairing on electron
Fermi surfaces in FeTe$_{0.55}$Se$_{0.45}$. $Phys.~Rev.~B$ \textbf{85},
094506 (2012).

\bibitem{Abdel-Hafiez} Abdel-Hafiez, M., He, Z., Zhao, J.,
Lu, X., Luo, H., Dai, P. and Chen, X.-J. Crossover
of the pairing symmetry from $s$- to $d$-wave in iron pnictide superconductors.
Preprint at http://arxiv.org/abs/1502.07130v1 (2015).

\bibitem{BiswasPRB} Biswas, P.K., Balakrishnan, G., Paul, D.M., Tomy, C.V., Lees, M.R. and Hillier, A.D.
Muon-spin-spectroscopy study of the penetration depth of FeTe$_{0.5}$Se$_{0.5}$.
$Phys.~Rev.~B$ 81, 092510 (2010). 

\bibitem{FlechterN} Fletcher, J. D., Serafin, A., Malone, L., 
Analytis, J. G., Chu, J.-H., Erickson, A. S., Fisher, I. R. and Carrington, A.
Evidence for a nodal-line superconducting state in LaFePO. $Phys.~Rev.~Lett.$ \textbf{102}, 147001 (2009).

\bibitem{HashimotoN} Hashimoto, K., Yamashita, M., Kasahara, S., 
Senshu, Y., Nakata, N., Tonegawa, S., Ikada, K., Serafin, A., Carrington, A.,
Terashima, T., Ikeda,  H., Shibauchi, T. and Matsuda, Y. Line nodes in
the energy gap of superconducting BaFe2(As$_{1-x}$P$_{x}$)$_{2}$
single crystals as seen via penetration depth and thermal conductivity.
$Phys.~Rev.~B$ \textbf{81}, 220501 (2010).

\bibitem{YamashitaN} Yamashita, M., Senshu, Y., Shibauchi, T., Kasahara, S.,
Hashimoto, K., Watanabe, D., Ikeda, H., Terashima, T., Vekhter, I., 
Vorontsov, A. B. and Matsuda, Y. Nodal gap structure of superconducting
BaFe2(As$_{1-x}$P$_{x}$)$_{2}$ from angle-resolved thermal conductivity
in a magnetic field. $Phys.~Rev.~B$ \textbf{84}, 060507 (2011).

\bibitem{NakaiN} Nakai, Y., Iye, T., Kitagawa, S., 
Ishida, K., Kasahara, S., Shibauchi, T., Matsuda, Y. and 
Terashima, T. $^{31}$P and $^{75}$As NMR evidence for a residual density
of states at zero energy in superconducting BaFe2(As$_{0.67}$P$_{0.33}$)$_{2}$.
$Phys.~Rev.~B$ \textbf{81}, 020503 (2010).

\bibitem{HashimotoK} Hashimoto, K., Kasahara, S., Katsumata, R., 
Mizukami, Y., Yamashita, M., Ikeda, H., Terashima, T., Carrington, A., 
Matsuda, Y. and Shibauchi, T. Nodeless vs nodal order parameters in LiFeAs
and LiFeP superconductors. $Phys.~Rev.~Lett.$ \textbf{108}, 047003 (2012).

\bibitem{DongN} Dong, J. K., Zhou, S. Y., Guan, T. Y., Zhang, H., 
Dai, Y. F., Qiu,  X., Wang, X. F., He, Y., Chen, X. H. and Li, S. Y. Quantum
criticality and nodal superconductivity in the FeAs-based superconductor
KFe$_{2}$As$_{2}$. $Phys.~Rev.~Lett.$ \textbf{104}, 087005 (2010).

\bibitem{QiuN} Qiu, X., Zhou, S. Y., Zhang, H., Pan, B. Y., Hong, X. C.,
Dai, Y. F., Eom, M.J., Kim, J. S. and Li, S. Y. Nodal superconductivity
in Ba(Fe$_{1-x}$Ru$_{x}$)$_{2}$As$_{2}$ induced by isovalent Ru
substitution. $Physical~Review~X$ \textbf{2}, 011010 (2012).

\bibitem{SongN} Song, C.-L., Wang, Y.-L., Cheng, P., Jiang, Y.-P.,
Li, W., Zhang, T., Li, Z., He, K., Wang, L., Jia, J.-F., 
Hung, H.-H., Wu, C., Ma, X., Chen, X. and Xue Q.-K. Direct observation
of nodes and twofold symmetry in FeSe superconductor. 
$Science$~\textbf{332}, 1410-1413 (2010).



\bibitem{ZhangN} Zhang, Y., Ye, Z. R., Ge, Q. Q., Chen, F., Jiang, J.,
Xu, M., Xie, B. P. and Feng, D. L. Nodal superconducting-gap structure
in ferropnictide superconductor BaFe$_{2}$(As$_{0.7}$P$_{0.3}$)$_{2}$.
$Nature~Physics$ \textbf{8}, 371-375 (2012).



\bibitem{Kuroki} Kuroki, K., Usui, H., Onari, S., Arita, R. and Aoki, H.
Pnictogen height as a possible switch between high-$T_{{\rm c}}$
nodeless and low-$T_{{\rm c}}$ nodal pairings in the iron-based superconductors.
$Phys.~Rev.~B$ \textbf{79}, 224511 (2009).

\bibitem{Graser10} Graser, S., Kemper, A. F., Maier, T. A., Cheng, H.-P.,
Hirschfeld, P. J. and Scalapino, D. J. $Phys.~Rev.~B$ \textbf{81}, 214503
(2010).

\bibitem{Maiti11} Maiti, S., Korshunov, M. M., Maier, T. A. , Hirschfeld, P. J.
and Chubukov, A. V. $Phys.~ Rev.~Lett.$ \textbf{107}, 147002 (2011).

\bibitem{Thomale11} Thomale, R., Platt, C., Hanke, W., Hu, J. and  
Bernevig, B.A. $Phys.~Rev.~Lett.$ \textbf{107}, 117001 (2011).

\bibitem{Khodas12} Khodas, M. and Chubukov, A. V. $Phys.~Rev.~Lett.$
\textbf{108}, 247003 (2012).



\bibitem{Fernandes13} Fernandes, R. M., and Millis, A. J. 
$Phys.~Rev.~Lett.$ \textbf{110}, 117004 (2013).

\bibitem{FernandezN} Kang, J., Kemper, A.F. and Fernandes, R. M. 
Manipulation of Gap Nodes by Uniaxial Strain in Iron-Based
Superconductors. $Phys.~Rev.~Lett.$ \textbf{113}, 217001 (2014).

\bibitem{raman_mode} Kretzschmar, F., Muschler, B., B\"{o}hm, T., Baum, A.,
Hackl, R., Wen, H.-H., Tsurkan, V., Deisenhofer, J. and Loidl, A. 
$Phys.~Rev.~Lett.$ \textbf{110}, 187002 (2013).

\bibitem{Bohm} B\"{o}hm, T. , Kemper, A.F., Moritz, B., Kretzschmar, F., 
Muschler, B., Eiter, H.-M., Hackl, R., Devereaux, T.P., Scalapino, D.J. and
Wen, H.-H. A balancing act: Evidence for a strong subdominant $d$-wave
pairing channel in Ba$_{0.6}$K$_{0.4}$Fe$_{2}$As$_{2}$. Preprint
at http://arxiv.org/abs/arXiv:1409.6815v1 (2014).

\bibitem{Taillefer} Tafti, F. F.,  Juneau-Fecteau, A.,  Delage, M. A.,
Cotret, S., Reid, J-Ph., Wang, A. F., Luo, X-G., Chen, X. H., Doiron-Leyraud, N.
and Taillefer, L. Sudden reversal in the pressure dependence of
$T_{c}$ in the iron-based superconductor KFe$_{2}$As$_{2}$. 
$Nat.~Phys.$ \textbf{9}, 349 (2013).

\bibitem{Sonier} Sonier, J.E., Brewer, J.H. and Kiefl, R.F. 
$Rev.~Mod.~Phys.$ \textbf{72}, 769 (2000).

\bibitem{Evtushinsky} Evtushinsky, D.V., Inosov, D.S., Zabolotnyy, V.B.,
Viazovska, M.S., Khasanov, R., Amato, A., Klauss, H.-H., Luetkens, H.,
Niedermayer, Ch., Sun, G.L., Hinkov, V., Lin, C.T., Varykhalov, A., 
Koitzsch, A., Knupfer, M., B\"{u}chner, B., Kordyuk, A.A. and Borisenko, S.V.
Momentum-resolved superconducting gap in the bulk of Ba$_{1-x}$K$_{x}$Fe$_{2}$As$_{2}$
from combined ARPES and ${\mu}$SR measurements. 
$New~J.~Phys.$ \textbf{11}, 055069 (2009).

\bibitem{Zabolotnyy} Zabolotnyy, V.B., Evtushinsky, D.V., Kordyuk, A.A.,
Inosov, D.S., Koitzsch, A., Boris, A.V., Sun, G.L., Lin, C.T., Knupfer, M.,
B\"{u}chner, B., Varykhalov, A., Follath, R. and Borisenko, S.V. (${\pi}$,
${\pi}$) electronic order in iron arsenide superconductors. 
$Nature$ \textbf{457}, 569 (2009).

\bibitem{Hashimoto-Science} Hashimoto, K., Cho, K., Shibauchi, T., 
Kasahara, S., Mizukami, Y., Katsumata, R., Tsuruhara, Y., Terashima, T., 
Ikeda, H., Tanatar, M. A., Kitano, H., Salovich, N., Giannetta, R.W., Walmsley, P.,
Carrington, A., Prozorov, R. and Matsuda, Y. A Sharp Peak of the Zero-Temperature
Penetration Depth at Optimal Composition in BaFe$_{2}$(As$_{1-x}$P$_{x}$)$_{2}$.
$Science$ \textbf{336}, 1554 (2012).

\bibitem{Hisrchfeld13} Wang, Y., Kreisel, A., Hirschfeld, P.~J. and
Mishra, V. $Phys.~Rev.~B$ \textbf{87}, 094504 (2013).

\bibitem{Sonier2011} Sonier, J.E., Huang, W., Kaiser, C.V., Cochrane, C.,
Pacradouni, V., Sabok-Sayr, S.A., Lumsden, M.D., Sales, B.C., McGuire, M.A.,
Sefat, A.S. and Mandrus, D. Magnetism and Disorder Effects on Muon
Spin Rotation Measurements of the Magnetic Penetration Depth in Iron-Arsenic
Superconductors. $Phys.~Rev.~Lett.$ \textbf{106}, 127002 (2011).

\bibitem{Brandt} Brandt, E.H. Flux distribution and penetration depth
measured by muon spin rotation in high-$T_{{\rm c}}$ superconductors.
$Phys.~Rev.~B$ \textbf{37}, 2349 (1988).

\bibitem{Bastian} Suter, A. and Wojek, B.M. $Physics~Procedia$ \textbf{30}, 69 (2012).\\
 The fitting of the $T$-dependence of the penetration depth with
${\alpha}$ model was performed using the additional library BMW developped
by B.M. Wojek.

\bibitem{Tinkham} Tinkham, M. Introduction to Superconductivity,
$Krieger~Publishing~Company$, $Malabar,~Florida$, 1975.

\bibitem{carrington} Carrington, A. and Manzano, F. Magnetic penetration
depth of MgB$_{2}$. $Physica~C$ \textbf{385}, 205 (2003).

\bibitem{Fang} Fang, M.H., Pham, H.M., Qian, B., Liu, T.J., Vehstedt, E.K.,
Liu, Y., Spinu, L. and Mao, Z.Q.  Superconductivity close to magnetic
instability in Fe(Se$_{1-x}$Te$_{x}$)$_{0.82}$. 
$Phys.~Rev.~B$ \textbf{78}, 224503 (2008).

\bibitem{padamsee} Padamsee, H., Neighbor, J.E. and Shiffman, C.A.
Quasiparticle Phenomenology for Thermodynamics of Strong-Coupling
Superconductors. $J.~Low~Temp.~Phys.$ \textbf{12}, 387 (1973).

\bibitem{khasanovalpha} Khasanov, R., Shengelaya, A., Maisuradze, A.,
La Mattina, F., Bussmann-Holder, A., Keller, H. and M\"{u}ller, K.~A. Experimental
Evidence for Two Gaps in the High-Temperature La$_{1.83}$Sr$_{0.17}$CuO$_{4}$
Superconductor. $Phys.~Rev.~Lett.$ \textbf{98}, 057007 (2007).

\bibitem{Fernandes11} Fernandes, R.~M. and Schmalian, J. 
$Phys.~Rev.~B$ \textbf{84}, 012505 (2011).

\bibitem{Stanev11} Stanev, V., Alexandrov, B. S., Nikolic, P. and 
Tesanovic, Z. $Phys.~Rev.~B$ \textbf{84}, 014505 (2011).

\bibitem{Fernandes2_2013}  Fernandes, R.~M. and Millis, A.~J. 
$Phys.~Rev.~Lett.$ \textbf{111}, 127001 (2013).

\bibitem{Hashimoto-pnictogen} Hashimoto, K., Kasahara, S., Katsumata, R.,
Mizukami, Y., Yamashita, M., Ikeda, H., Terashima, T., Carrington, A.,
Matsuda, Y. and Shibauchi, T. Nodal versus Nodeless Behaviors of the
Order Parameters of LiFeP and LiFeAs Superconductors from Magnetic
Penetration-Depth Measurements. 
$Phys.~Rev.~Lett.$ \textbf{108}, 047003 (2012).



\bibitem{Walsmley} Walmsley, P., Putzke, C., Malone, L., Guillamon, I.,
Vignolles, D., Proust, C., Badoux, S., Coldea, A.I., Watson, M.D., Kasahara, S.,
Mizukami, Y., Shibauchi, T., Matsuda, Y. and Carrington, A. Quasiparticle
Mass Enhancement Close to the Quantum Critical Point in BaFe$_{2}$(As$_{1-x}$P$_{x}$)$_{2}$.
$Phys.~Rev.~Lett.$ \textbf{110}, 257002 (2013).

\bibitem{Levchenko13} Levchenko,  A., Vavilov,  M.G., Khodas,  M. and
Chubukov, A.V. $Phys.~Rev.~Lett.$ \textbf{110}, 177003 (2013).

\bibitem{Sachdev} Chowdhury, D., Swingle, B., Berg, E. and
Sachdev, S. Singularity of the London Penetration Depth at Quantum
Critical Points in Superconductors. 
$Phys.~Rev.~Lett.$ \textbf{111}, 157004 (2013).

\bibitem{Giacomo} Prando, G., Hartmann, Th., Schottenhamel, W., Guguchia, Z.,
Sanna, S., Ahn, F., Nekrasov, I., Wolter, A.~U.~B., Wurmehl, S., Khasanov, R.,
Eremin, I. and B\"{u}chner, B. Mutual independence of critical temperature
and superfluid density under pressure in optimally electron-doped
superconducting LaFeAsO$_{1-x}$F$_{x}$. Preprint at http://arxiv.org/abs/arXiv:1502.02713
(2015).



\bibitem{Maisuradze} Maisuradze, A., Graneli, B., Guguchia, Z., Shengelaya, A.,
Pomjakushina, E., Conder, K. and Keller, H. Effect of pressure on the
Cu and Pr magnetism in Nd$_{1-x}$Pr$_{x}$Ba$_{2}$Cu$_{3}$O$_{7}$
investigated by muon spin rotation. $Phys.~Rev.~B$ \textbf{87}, 054401
(2013).

\bibitem{Andreica} Andreica, D. 2001 $Ph.D.~thesis$ IPP/ETH-Z\"{u}rich.

\bibitem{Maisuradze-PC}  Maisuradze, A., Shengelaya, A., Amato, A., 
Pomjakushina, E. and Keller, H. Muon spin rotation investigation of the
pressure effect on the magnetic penetration depth in YBa$_{2}$Cu$_{3}$O$_{x}$.
$Phys.~Rev.~B$ \textbf{84}, 184523 (2011).\end{thebibliography}

\begin{thebibliography}{10}
\bibitem{Budko} Sergey L. Bud'ko, Ni Ni, and Paul C. Canfield, Phys.
Rev. B 79, 220516(R) (2009).



\bibitem{Suter} A. Suter and B.M. Wojek, $Physics~Procedia$ \textbf{30},
69-73 (2012).

\bibitem{Huang} Q.~Huang, Y.~Qiu, W.~Bao, M.A.~Green, J.W.~Lynn,
Y.C.~Gasparovic, T.~Wu, G.~Wu, X.H.~ChenXH, Phys. Rev. Lett. \textbf{101},
257003 (2008).

\bibitem{Torikachvili} M.S. Torikachvili, S.L. Bud'ko, N. Ni, and
P.C. Canfield, Phys. Rev. Lett. \textbf{101}, 057006 (2008).

\bibitem{Miclea} C.F. Miclea, M. Nicklas, H.S. Jeevan, D. Kasinathan,
Z. Hossain, H. Rosner, P. Gegenwart, C. Geibel, and F. Steglich, Phys.
Rev. B \textbf{79}, 212509 (2009).

\bibitem{Zhao} J. Zhao, Q. Huang, C. de la Cruz, S. Li, J.W. Lynn,
Y. Chen, M.A. Green, G.F. Chen, G. Li, Z. Li, J.L. Luo, N.L. Wang,
and P. Dai, Nature Materials \textbf{7}, 953 (2008).

\bibitem{Khasanov-2009} R. Khasanov, D.V. Evtushinsky, A. Amato, H.-H.
Klauss, H. Luetkens, Ch. Niedermayer, B. B\"{u}chner, G.L. Sun, C.T. Lin,
J.T. Park, D.S. Inosov, and V. Hinkov. Phys. Rev. Lett. \textbf{102},
187005 (2009).

\bibitem{Kubo} R.~Kubo and T.~Toyabe, \textit{Magnetic Resonance
and Relaxation} (North Holland, Amsterdam, 1967).

\bibitem{Hayano} R.S. Hayano, Y.J. Uemura, J. Imazato, N. Nishida,
T. Yamazaki, and R. Kubo, Phys. Rev. B \textbf{20}, 850 (1979).

\bibitem{Kang14}  Jian Kang, Alexander F. Kemper, and Rafael M. Fernandes.
Phys. Rev. Lett. \textbf{113}, 217001 (2014).\end{thebibliography}
\end{document}